\documentclass[a4paper,twocolumn,english,aps,floatfix]{revtex4}
\usepackage{epsfig,amsmath}
\begin{document}

\title{Spin-based all-optical quantum computation with quantum dots:\\ understanding and suppressing decoherence}
\author{T. Calarco$^1$, A. Datta$^2$, P. Fedichev$^3$, E. Pazy$^4$, P. Zoller$^3$}

\affiliation{$^1$NIST, Gaithersburg, MD 20899-8423, USA, and ECT*, I-38050 Villazzano (TN), Italy\\
$^2$Department of Electrical Engineering, Indian Institute of Technology, Kanpur 208016, India \\
$^3$Institut f\"ur Theoretische Physik, Universit\"at Innsbruck,
A-6020 Innsbruck, Austria\\
$^4$Department of Physics, Ben-Gurion University of the Negev,
Beer-Sheva 84105, Israel }

\begin{abstract}
We present an all-optical implementation of quantum computation
using semiconductor quantum dots. Quantum memory is represented by
the spin of an excess electron stored in each dot. Two-qubit gates
are realized by switching on trion-trion interactions between
different dots. State selectivity is achieved via conditional
laser excitation exploiting Pauli exclusion principle. Read-out is
performed via a quantum-jump technique. We analyze the effect on
our scheme's performance of the main imperfections present in real
quantum dots: exciton decay, hole mixing and phonon decoherence.
We introduce an adiabatic gate procedure that allows one to
circumvent these effects, and evaluate quantitatively its
fidelity.
\end{abstract}
\maketitle

\section{Introduction}
The promise of quantum computation is to enable algorithms which
render feasible problems requiring exorbitant resources for their
solution on a classical computer. This has stimulated a large
number of proposals for the physical implementation of the
elementary logical operations building up a general-purpose
quantum computer \cite{FdP}. A central issue is the trade-off
between efficient coupling to the system, in order to control the
quantized degrees of freedom, and good isolation from the
environment, in order to preserve the coherence of the quantum
evolution. Strategies have been developed to fight decoherence
taking place during the computation, ranging from ``active''
(error-correcting) to ``passive'' (error-avoiding) schemes.
Thereby, unwanted physical processes (i.e., computational errors)
of a {\it general} kind can be compensated for, either by
detecting and correcting their effect via redundant qubit encoding
\cite{Error-correcting}, or by decoupling the qubits from the
environment dynamics through algebraic techniques exploiting
symmetries in the evolution \cite{Error-avoiding}. A less general,
more implementation-dependent approach is to study the specific
decoherence channels of a certain physical scheme, and to design
gating processes that are stable against the relevant types of
errors (for an example with ion traps, see Ref.~\cite{ionPRA}).

In this paper, we adopt the latter point of view, and apply it to
a recent proposal for all-optical quantum information processing
based on charged semiconductor quantum dots \cite{pauli}. In this
scheme, quantum information is stored in the spin of an excess
electron in a quantum dot (QD), and gating between 2 QDs is
performed via optical excitation of electron-hole pairs
(excitons), which in an external electric field acquire a dipole
moment allowing them to interact with each other. In this way, the
quantum memory coherence time is in the $\mu$s range, typical for
spin degrees of freedom in semiconductor heterostructures
\cite{spindeph}, while the two-qubit gating time is in the ps
range, as dictated by the electrostatic dipole-dipole interaction.

Decoherence is important mainly during gate operation when
excitonic states are created that interact with the phonon bath.
Recent calculations for the strong field limit \cite{kuhn}
indicate that the leading dephasing mechanism is the coupling to
acoustic phonons. In this paper, we focus on this coupling
mechanism, and we describe a procedure that allows one to
circumvent, to a large extent, the limitations imposed by this
decoherence channel on the fidelity $F$ of the gate operation. We
will evaluate the dependence of $F$ on various parameters,
including temperature, and take into account other sources of
imperfection like heavy-/light-hole mixing.

We choose to neglect all kinds of nonidealities arising from
limitations, e.g., in QD fabrication and manipulation techniques.
We are well aware that these might yield the most significant
problems for the implementation of our proposal in the immediate
future. However, we prefer to focus on fundamental
quantum-mechanical limitations of our physical system rather than
on technical problems. Once the technological advances have
overcome the latter, the relevant part will be to find ways to
circumvent the former. The main purpose of this paper is to
develop strategies aimed at this.

The paper is organized as follows: in Sec. II we describe the
general idea of a two-qubit quantum gate based on selective
switching of controlled interactions. In Sec. III we recall the
dynamics of charge carriers in a quantum dot, including external
static and oscillating electromagnetic fields. In Sec. IV we
derive few-level model corresponding to the above general scheme
and discuss some of its limitations. In Sec. V we discuss our
two-qubit gate and develop its adiabatic version, suitable for
operation even in realistic scenarios with hole mixing. In Sec. VI
we propose a hole-mixing tolerant scheme for single-qubit
operations. In Sec. VII we analyze the effect of the interaction
with phonons on the performance of our adiabatic gates, showing
that the gates indeed are quite robust also against this kind of
imperfection. In Sec. VIII we describe how the quantum jump
technique can be employed for measuring the spin state of a
confined electron, emphasizing that this can be done even for the
case of non-zero hole mixing. Our conclusions are summarized in
Sec. IX.

\section{Quantum gate model}

To execute an arbitrary quantum computation, i.e. to control the
coherent evolution of a system composed of an arbitrary number of
qubits, one does not need to realize physically arbitrary
multi-qubit operations. On the contrary, just two kinds of
elementary operations are sufficient, out of which all others can
be constructed. These two elementary gates are the set of
rotations of a single qubit, and a specific entangling operation
on two qubits. Among the possible choices for the latter, one
which is well suited for an implementation with atomic-like
systems like quantum dots is the phase gate -- a transformation
which rotates by a certain phase just one component of logical
states:
\begin{equation}
\begin{array}{rcrl}
|0\rangle|0\rangle &\longrightarrow &|0\rangle|0\rangle&\\
|0\rangle|1\rangle &\longrightarrow &|0\rangle|1\rangle&\\
|1\rangle|0\rangle &\longrightarrow &|1\rangle|0\rangle&\\
|1\rangle|1\rangle &\longrightarrow
&e^{i\vartheta}|1\rangle|1\rangle &\!\!\!.
\end{array}
\label{defGate}
\end{equation}
When we have $\vartheta=\pi$, this is equivalent, up to
single-qubit rotations, to a controlled-NOT gate. Ideally, this
would be accomplished by means of a state-dependent interaction of
the form
\begin{equation}
\label{Hideal} H_{\mathrm{ideal}}=\Delta E_{ab}(t)\left|
1\right\rangle _{a}\left\langle 1\right| \otimes \left|
1\right\rangle _{b}\left\langle 1\right|.
\end{equation}
This describes a situation in which the two-qubit system undergoes
an energy shift $\Delta E_{ab}$ if and only if both qubits are in
state $|1\rangle$. Imposing the additional condition
\begin{equation}
\int_{t_{0}}^{t_{0}+\tau }\Delta E_{ab}(t^{\prime })dt^{\prime
}=\vartheta
\end{equation}
on the time dependence of the energy shift, Eq.~(\ref{defGate}) is
recovered.

\subsection{Phase gate model: auxiliary interacting states}
\begin{figure}[h]
\begin{center}
\epsfig{figure=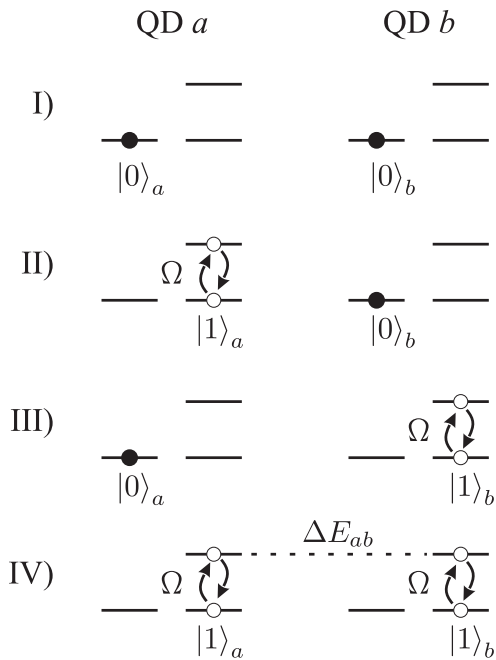,width=6.5truecm}
\caption{\label{gateop}Gate operation via an auxiliary state: in
the ideal scenario when only one of the logical states can be
coupled to the interacting state, the interaction leading to the
logical phase is ``switched on'' only when both qubits are in the
same state (here, $|1\rangle$).}
\end{center}
\end{figure}

An interaction of the form Eq.~(\ref{Hideal}) is not
straightforwardly found in nature. Implementing it entails of
course a certain degree of engineering ``natural'' interactions,
i.e., those directly available in a specific physical system.
This, together with other requirements on the stability of the
available quantum memory, affects the choice of the particular
qubit implementation. When it comes to systems of confined
electrons in solid state systems, such as quantum dots, two
different choices are natural for the logical degree of freedom:
either charge excitation \cite{RossiPRL}, or spin polarization
\cite{LossDiVincenzo}. The former provides for a strong
interaction, leading to comparatively shorter gate times but to
faster decoherence rates as well; conversely, the latter suffers
less from the coupling to the environment, yielding better
stability against memory decoherence, but bears also a weaker
coupling between qubits, requiring longer times for gate
operation. Aiming at a high ratio between coherence time and gate
operation time leads to conflicting requirements. Reasonable
trade-offs can be achieved in each case -- however, sticking to
the same degree of freedom for both the memory and the two-qubit
interaction may  be not necessarily the only option. For instance,
the same effect of the interaction Eq.~(\ref{Hideal}) can be
obtained by introducing an auxiliary state $|x\rangle$. Let us
consider two qubits, labeled by $a$ and $b$ and with logical
states $|\alpha\rangle_{a,b}$ ($\alpha\in\{0,1\}$). Each qubit is
selectively coupled to a further state $|x\rangle$ -- namely, only
$|1\rangle$ can be excited to $|x\rangle$. This situation is
described by the following Hamiltonian:
\begin{eqnarray}
H_{\mathrm{phys}}(t)&=&\!\sum_{\substack{\alpha =0,1,x\\
\nu=a,b}}\!\!E_{\alpha }\left| \alpha \right\rangle _{\nu
}\left\langle \alpha \right|+\frac{\Omega (t)}{2}\sum_{\nu
=a,b}\left|
x\right\rangle _{\nu }\left\langle 1\right| +\mathrm{h.c.}\nonumber\\
&&\mbox{}+\Delta E_{ab}(t)\left| x\right\rangle _{a}\left\langle
x\right| \otimes \left| x\right\rangle _{b}\left\langle x\right|
 \label{Hphys}
\end{eqnarray}
As in Ref.~\cite{pauli}, the logical states $|0\rangle$ and
$|1\rangle$ (the quantum memory) can be encoded into
long-coherence spin states, while the auxiliary states
$|x\rangle$, needed for the gate to be performed, can be chosen to
be electrostatically interacting states. State selectivity,
required for conditional logical operations, is accomplished via
the state-dependent coupling $\Omega (t)$. The simplest strategy
for performing a quantum gate exploiting the coupling scheme
Eq.~(\ref{Hphys}) would be, e.g., to selectively excite the
interacting state $|x\rangle$ via a Rabi flop, wait for the
desired gate phase to be accumulated, and then de-excite. The
interaction energy shift would then be effective only if both QDs
started off state $|1\rangle$, as described in Fig.~\ref{gateop}.
This procedure works in the ideal case when the coupling is
perfectly state-selective as in Eq.~(\ref{Hphys}). Compared to
similar schemes for neutral atoms (see e.g. \cite{Rydberg}), it
has also the advantage that quantum dots, unlike trapped atoms,
are not subject to back-action on motional dynamics. However, in a
real situation state selectivity may not be perfectly satisfied,
in which case the simple procedure described above would not work.
We will handle this imperfection below, and develop a strategy for
overcoming it. But first we need a model of quantum dot dynamics
that can account for Eq.~(\ref{Hphys}). This is the subject of
Sect.~\ref{sec:QDdynamics}.

\subsection{Gate fidelity}
To evaluate the performance of a quantum gate, one needs to
compare its desired operation, Eq.~(\ref{defGate}) in our case,
with the actual performance of the physical system which
implements it. The fidelity $F$ represents a quantitative basis
for this. To define it, let us start from the logical input state
\begin{equation} \left| \chi \right\rangle
\equiv c_{00}\left| 00\right\rangle +c_{01}\left| 01\right\rangle
+c_{10}\left| 10\right\rangle +c_{11}\left| 11\right\rangle
=\sum_{n=0}^{3}c_{n}\left| n\right\rangle,
\end{equation}
which is an arbitrary superposition of all two-qubit computational
basis states. The goal of gate operation is to obtain the ideal
output
\begin{equation} \left| \tilde{\chi}\right\rangle
\equiv\sum_{n=0}^{3}e^{i\phi _{n}}c_{n}\left| n\right\rangle,
\label{chitilde}
\end{equation}
This is equivalent to the desired two-qubit transformation
Eq.~(\ref{defGate}): one can be recovered from the other by
redefining the logical states via single-qubit operations. The
gate phase $\vartheta$ turns out to be related to the logical
phases $\phi_{\alpha\beta}$ as follows \cite{ionPRA}:
\begin{equation}
\vartheta = \phi _{00}-\phi _{01}-\phi _{10}+\phi _{11}.
\end{equation}
Thus the condition $\vartheta\stackrel{!}=\pi$ simply translates
into a condition on the $\phi_{\alpha\beta}$'s.

The actual physical situation may involve other external (i.e.,
non-logical) degrees of freedom, which are not perfectly under
control. In this case the initial state $\sigma$ will rather be a
mixture:
\begin{equation} \label{initialsigma}\sigma  \equiv \left| \chi
\right\rangle \left\langle \chi \right| \otimes \rho _{\rm
ext}(t_{0}),
\end{equation}
where $\rho_{\rm ext}$ denotes the density matrix for external
degrees of freedom. The operation $\cal U$ realized in the lab
will in general involve both internal and external degrees of
freedom in a non-trivial way: therefore the actual output
\begin{equation}
\sigma ^{\prime } =\mathcal{U}\sigma \mathcal{U}^{\dagger }
\end{equation}
will no longer be written in a simple factorized form like
Eq.~(\ref{initialsigma}). In order to compare this state with the
ideal one which would be obtained in the case of perfect
operation, we define the fidelity
\begin{equation} F =\min_{\chi
}\mathrm{tr}_{\rm ext}\left\langle \tilde{\chi}\right| \sigma
^{\prime }\left| \tilde{\chi}\right\rangle.
\end{equation}
The intuitive meaning of this definition is that of a worst-case
estimate (hence the minimum over the possible inputs
$|\chi\rangle$) of the gate performance, averaged over the
available non-logical states not being under control (hence the
trace over the external degrees of freedom). Another option would
be to define the fidelity as an average over the logical inputs
$|\chi\rangle$ -- however, we will stick to the minimum fidelity
which gives a lower bound to the average fidelity.

Taking the most general possible form for the external state,
\begin{equation}
\rho _{ext} =\sum_{E}p_{E}\left| E\right\rangle \left\langle
E\right|,
\end{equation}
and assuming that the evolution does not mix different logical
states,
\begin{equation} \mathcal{U}\left| \chi
\right\rangle \left| E\right\rangle \approx
\sum_{n=0}^{3}c_{n}\left| n\right\rangle \otimes V_{n}\left|
E\right\rangle,
\end{equation}
the fidelity takes the form
\begin{equation}
F =\min_{\{c_{n}\}}\sum_{m,n=0}^{3}\left| c_{m}\right| ^{2}\left|
c_{n}\right| ^{2}e^{i(\phi _{n}-\phi _{m})}\mathrm{tr}_{ext}\left(
V_{m}V_{n}^{\dagger }\right),
\end{equation}
which will be relevant for our calculations below.

\section{\label{sec:QDdynamics}Quantum Dot dynamics}
Quantum dots, due to their discrete density of states, are a very
promising candidate for the implementation of quantum information
processing \cite{RossiPRL,Qdimplem}. The brilliant idea first
proposed by DiVincenzo and Loss \cite{LossDiVincenzo} to employ
the spin of an electron confined in a QD as the qubit degree of
freedom has been developed by the authors over the years
\cite{Lossrecent} and is now pursued by many groups
\cite{spinqu,Imamoglu00,QED}. Combining QD technology with
ultra-fast laser pulses now seems to be one of the most promising
channels for such an implementation scheme \cite{pauli,spinoptic}.
Recently the necessary coherence required for such a task, i.e.,
Rabi oscillations, has been experimentally observed \cite{Rabi}.
There have been also impressive experimental achievements in
exciting and probing excitons in QDs \cite{qdexciton}.

The complex many-body dynamics of charge carriers in a
semiconductor can be considerably simplified when considering
semiconductor heterostructures like quantum wells and dots. The
purpose of the present Section is to write down explicitly the
carrier Hamiltonian for a quantum dot under these approximations.
In the next Section this description will be linked to the
particular model described by Eq.~(\ref{Hphys}). Two main
approximations \cite{Cardona,Hawrylak} are understood throughout
the following. The first is the effective mass approximation,
which arises from approximating the band dispersion relation
around a band extremum up to second order in the carrier
wavevector $k$. This is valid for small values of $k$, and allows
for simplifying Hamiltonians in terms of effective electron and
hole masses that take into account the underlying many-body
dynamics. The other is the envelope function approximation
\cite{Rossi}, which is based on the following assumptions: (i) the
different materials constituting the heterostructure are perfectly
lattice-matched; (ii) the periodic parts of the Bloch functions
$u_{l,\mathbf{0}}(\mathbf{r})$ are the same in the different
layers; (iii) the confining potential is smoothly varying on the
scale of the lattice structure, apart possibly from abrupt
interfaces. The wavefunction can then be expanded as a sum of
products of the rapidly varying functions
$u_{l,\mathbf{0}}(\mathbf{r})$ by slowly varying envelope
functions which obey an effective Schr\"{o}dinger equation
involving the effective masses. We consider QDs in the
``strong-confinement'' regime, in which the typical length scale
in the growth direction $L$ is of the order of 10~nm to 20~nm.
Considering QDs in the strong-confinement regime means that all
relevant energy scales, e.g. charge carrier interactions in the QD
or electron phonon interactions, will be small compared to the
level spacing of the QD, typically of the order of $25$ meV for
electrons.

\subsection{Single-particle states under external
fields}

Under the above approximations, the carrier Hamiltonian for a
quantum dot can be written as \cite{Hawrylak,RossiPRB}
\begin{equation}
H^{c}=H_{\perp }^{c}+H_{\parallel }^{c}=H_{\perp }^{e}+H_{\perp
}^{h}+H_{\parallel }^{e}+H_{\parallel }^{h}
\end{equation}
The electron in-plane Hamiltonian $H_{\parallel }^{e}$ describes
the confinement in the direction perpendicular to the QD symmetry
axis $\hat z$, which can be modeled with a parabolic potential:
\begin{equation}
H_{\parallel }^{e}=-\frac{\hbar ^{2}}{2m_{e}}\nabla _{\mathbf{r}}^{2}+\frac{m_{e}\omega _{e}^{2}}{2}r^{2}+e%
\mathbf{F}\cdot \mathbf{r}
\end{equation}
where the in-plane coordinate vector is $\mathbf{r}\equiv(x,y)$,
and the electrical field $\mathbf{F}$ is taken to be parallel to
the $xy$ plane. Defining
\begin{equation}
\mathbf{r}_{e}=\mathbf{r}+\frac{e\mathbf{F}}%
{m_{e}\omega _{e}^2}=(r_{e},\theta _{e}),
\end{equation}
the eigenstates $\left| n,q\right\rangle ^{e}$ of $H_{\parallel
}^{e}$ in coordinate representation are
\begin{equation}
\left\langle r_{e}|n,q\right\rangle ^{e}=\frac{r_{e}^{|q|}\sqrt{n_{r}!}%
e^{iq\theta _{e}-r_{e}^{2}/(2l_{0})^{2}}}{l_{0}^{|q|+1}\sqrt{\pi (n_{r}+|q|)!%
}}\mathcal{L}_{n_{r}}^{|q|}\left(
\frac{r_{e}^{2}}{l_{0}^{2}}\right)
\end{equation}
where $n=0,1,\ldots $ is the principal,
$q=-n,-n+2,\ldots ,n-2,n$ the azimuthal, and $n_{r}=(n-|q|)/2$ the radial quantum number; $\mathcal{%
L}_{n_{r}}^{|q|}\left( z\right) $ are Laguerre polynomials;
\begin{equation}
l_{0}=\sqrt{\frac{\hbar }{2m_e\omega _{0}}}
\end{equation}
and the eigenenergies are
\begin{equation}
\epsilon _{nq}^{e}=\hbar (n+1)\omega _{0}.
\end{equation}
In the growth direction the perpendicular Hamiltonian $H_{\perp
}^{e}$ is in general a very narrow potential given by the quantum
well structure and is therefore typically approximated by a step
like potential. In the strong-confinement regime, a good
approximation is to assume that the system remains in its ground
state. Thus the problem effectively reduces to the in-plane
dynamics. The hole Hamiltonian $H_{\perp ,\parallel }^{h}$ is of
course the same as $H_{\perp ,\parallel }^{e}$ but with hole
parameters $m_{h}$ and $\omega _{h}$, and opposite charge.

Due to the strong confinement and spatially symmetric shapes of
the confining potentials in QDs, electronic angular momentum
states can be defined. They exhibit many atomic-like symmetries
which have been experimentally identified \cite{atomlike}. In
contrast to atoms, when considering the quantum numbers defining
the angular momentum of an electron or a hole confined in a QDs,
one has to take into account the underlying band structure
\cite{Bastard}.

Taking spin into account leads to splitting into hole subbands.
The valence band, built from atomic $p$-type orbitals, contains
states carrying an internal (band) angular momentum $\mathbf{m}$
equal to unity. Thus the total angular momentum is
\begin{equation}
\mathbf{j}=\mathbf{\sigma }+\mathbf{m}+\mathbf{l}
\end{equation}
where $\mathbf{\sigma }$ is the spin and $\mathbf{l}$ the orbital
angular momentum. Good quantum numbers are the modulus of
$\mathbf{j}$ and its component along the QD\ symmetry axis
$\hat{z}$. The single-particle states of the valence band with
$\mathbf{l}=0
$ are classified according to the value of $(|\mathbf{\sigma }+\mathbf{m}|,\sigma _{z}+m_{z})$%
, as follows:
\begin{description}
\item  $(3/2,\pm 3/2)$: heavy-hole subband;
\item  $(3/2,\pm
1/2)$: light-hole subband;
\item  $(1/2,\pm 1/2)$: spin-orbit
split-off subband.
\end{description}

For the dynamics considered in this paper, only heavy and light
holes will matter, the split-off subband being energetically far
apart. So let us define electron and hole operators for the QD labeled by $\nu$ ($\nu \in \{a,b\}$), with composite index $%
i=[n,q]$ and spin $\sigma $:
\begin{eqnarray}
c_{\nu ,i,\sigma }^{\dagger }\left| \mathrm{vac}\right\rangle
&=&\left|
i,\sigma \right\rangle _{\nu }, \\
h_{\nu ,j,\sigma ^{\prime }}^{\dagger }\left|
\mathrm{vac}\right\rangle &=&\left| j,\sigma ^{\prime
}\right\rangle _{\nu }.
\end{eqnarray}
We can now write the noninteracting part of the carrier
Hamiltonian for the QD $\nu $:
\begin{equation}
H_{\nu }^{c}=\sum_{i,\sigma =\pm 1/2}\epsilon _{i,\sigma
}^{e}c_{\nu ,i,\sigma }^{\dagger }c_{\nu ,i,\sigma
}+\sum_{j;\sigma ^{\prime }=-3/2}^{3/2}\epsilon _{j,\sigma
}^{h}h_{\nu ,j,\sigma ^{\prime }}^{\dagger }h_{\nu ,j,\sigma
^{\prime }}
\end{equation}

\subsection{Carrier-carrier interaction}
The electrostatic interaction Hamiltonian is written as
\begin{eqnarray}
H_{\nu }^{cc} &=&\sum_{\substack{i,j,k,l \\ \sigma ,\sigma ^{\prime }}}%
\frac{1}{2}\left( \left\langle ij|V|kl\right\rangle _{ee}c_{\nu
,i,\sigma }^{\dagger }c_{\nu ,j,\sigma ^{\prime }}^{\dagger
}c_{\nu ,l,\sigma }c_{\nu
,k,\sigma ^{\prime }}\right.   \nonumber \\
&&\left. +\left\langle ij|V|kl\right\rangle _{hh}h_{\nu ,i,\sigma
}^{\dagger }h_{\nu ,j,\sigma ^{\prime }}^{\dagger }h_{\nu
,l,\sigma }h_{\nu ,k,\sigma
^{\prime }}\right)  \\
&&-\left\langle ij|V|kl\right\rangle _{eh}c_{\nu ,i,\sigma
}^{\dagger }h_{\nu ,j,\sigma ^{\prime }}^{\dagger }c_{\nu
,l,\sigma }h_{\nu ,k,\sigma ^{\prime }}  \nonumber
\end{eqnarray}
where the matrix elements of the Coulomb potential
\begin{equation}
V(\mathbf{r}-\mathbf{r}^{\prime })=\frac{e^{2}}{4\pi\epsilon\left|
\mathbf{r}-\mathbf{r}^{\prime }\right| }
\end{equation}
are calculated on electron and/or hole wavefunctions, according to
subscripts. Here, carrier number conservation is assumed, since
processes violating this (like e.g. Auger recombination and impact
ionization) are relevant for energies and densities higher than we
are considering \cite{RossiPRB}. It should be noted that, in
contrast to higher dimensional quantum structures, in QDs
carrier-carrier interactions only induce an energy level
renormalization without causing scattering or dephasing.

\subsection{Interaction with a laser field}
Let us consider a laser of amplitude $E(t)$ and central frequency $%
\omega _{L}$ impinging on our QD. Under the dipole and
rotating-wave approximations \cite{dipoleapp} the corresponding
Hamiltonian is
\begin{equation}
H_{\nu }^{\mathrm{int}}=-\sum_{i,j,\sigma ,\sigma ^{\prime
}}\left[ \mu _{ij}^{\sigma \sigma ^{\prime }}E^{\ast
}(t)e^{-i\omega _{L}t}c_{\nu
,i,\sigma }^{\dagger }h_{\nu ,j,\sigma ^{\prime }}^{\dagger }+\mathrm{h.c.}%
\right],   \label{Hint}
\end{equation}
where $\mu _{ij}^{\sigma \sigma ^{\prime }}$ is the dipole matrix
element between the wave functions of an electron with spin
$\sigma $ in state $i$ and a hole having angular momentum with $z$
component $\sigma ^{\prime }$ in
state $j$. The resulting selection rule is that the change in the number of electron-hole pairs  can only be $%
\Delta N=\pm 1$.

\section{Three-level model\label{sec:threelevel}}
Let us consider the QD
identified by the index $\nu $, with an excess electron in the
conduction-band ground state. We label its spin states with
\begin{eqnarray}
\left| 0\right\rangle _{\nu } &\equiv &c_{\nu ,0,-1/2}^{\dagger
}\left|
\mathrm{vac}\right\rangle,  \label{def0}\\
\left| 1\right\rangle _{\nu } &\equiv &c_{\nu ,0,1/2}^{\dagger
}\left| \mathrm{vac}\right\rangle.\label{def1}
\end{eqnarray}
These are eigenstates of the bare Hamiltonian $H_\nu^c$, with
eigenvalues $\epsilon^e_{0,-1/2}$ and $\epsilon^e_{0,1/2}$
respectively, and are not affected by the carrier-carrier
interaction $H^{cc}_\nu$. On the other hand, in the so-called
``trion'' --~i.e., the state obtained from
Eqs.(\ref{def0}-\ref{def1}) by creating an exciton, having
``bare'' energy
$\epsilon^e_{0,1/2}+\epsilon^e_{0,-1/2}+\epsilon^h_{0,\sigma_h}$~--,
more than one carrier is present in the QD. Therefore, in this
case the interaction $H^{cc}_\nu$ changes the bare state $c_{\nu
,0,+1/2}^{\dagger
}c_{\nu ,0,-1/2}^{\dagger }h_{\nu ,0,\sigma _{h}}^{\dagger }\left| \mathrm{%
vac}\right\rangle$ into the physical interacting state $\left|
x,\sigma _{h}\right\rangle _{\nu }$, which we will take as our
auxiliary state for gate operation. Such states were observed and
studied experimentally in single self-assembled QDs \cite{trions}.

According with the above selection rule, a laser pulse can excite
at most one exciton. If the laser is tuned on the lowest interband
excitation energy, then a ground-state exciton can be obtained.
Due to angular momentum conservation, the hole angular momentum
$\sigma _{h}$ will depend upon the laser polarization. For
instance, in the case of a semiconductor material where heavy
holes have the lowest energy, and assuming $\sigma ^{+}$
circularly polarized light, the only hole state that can be
excited has $\sigma _{h}=3/2$. If moreover the absolute value of
the Rabi frequency, defined as
\begin{equation}
\Omega (t)\equiv \frac{2\mu _{00}^{-1/2,+3/2}E(t)}{\hbar },
\end{equation}
is much smaller than the intraband excitation energy, $\left|
\Omega \right| \ll \omega _{e,h}$, then we can neglect the
probability of promoting the electron from the valence band to a
higher-excited conduction band state. Under these assumptions, the
interaction Hamiltonian Eq. (\ref{Hint}) simplifies to
\begin{equation}
H_{\nu }^{\mathrm{int}}=\hbar \Omega (t)e^{-i\omega _{L}t}c_{\nu
,0,-1/2}^{\dagger }h_{\nu ,j,+3/2}^{\dagger }
\end{equation}
If the temperature is sufficiently low with respect to the
electronic intraband excitation energy, $k_{B}T\ll \hbar \omega
_{e}$, then we can neglect also the excited states of the excess
electron.

In the subspace defined by the states $\{\left| 0\right\rangle
_{\nu },\left| 1\right\rangle _{\nu },\left| x,+3/2\right\rangle
_{\nu }\}$, the effect of the carrier-carrier interaction term
$H_{\nu }^{cc}$ will be to change the energy of the trion states.
In particular, an external static electric field $\mathbf F$
applied in the $(x,y)$ plane will mutually displace the
wavefunctions of the electron and of the hole which constitute an
exciton, since they have opposite charge. In this way, the trion
states acquire an electric dipole moment. The electrostatic
interaction will then shift the energy of a trion state where a
trion is also present in a neighboring dot. This energy
difference, the so-called trion-trion shift $\Delta E_{ab}$, will
be very important for obtaining the state-dependent phase needed
for the logical gate to be performed. The key ingredient for this
is a state-selective coupling of the logical states $\left|
0\right\rangle $ and $\left| 1\right\rangle $ to the auxiliary
interacting state $\left| x^{\uparrow }\right\rangle \equiv \left|
x,+3/2\right\rangle $. In the simplest, ideal case, this
state-selectivity can be obtained if only one of the logical
states is coupled to $\left| x^{\uparrow }\right\rangle $. The
possibility of realizing physically such an effect in QDs is
offered by the mechanism described in the following section.

\subsection{Exciton Pauli blocking}
\begin{figure}[h]
\begin{center}
\epsfig{figure=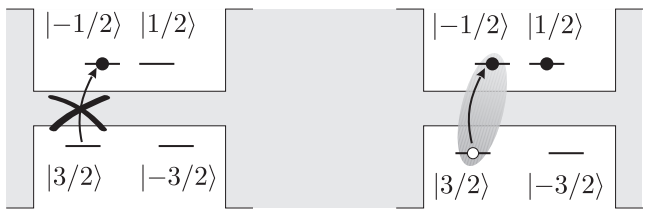,width=8.5truecm}
\caption{\label{Pauliblocking}Pauli-blocking mechanism: a pulse of
$\sigma_+$-polarized light can promote an electron from the
valence band to the conduction-band $-1/2$-spin state of a quantum
dot only if the latter is not occupied, i.e., if the excess
electron in the dot is in the opposite spin state (right).
Otherwise, no excitation takes place (left).}
\end{center}
\end{figure}

In fact, the Pauli exclusion principle forbids double occupancy of
any of the electronic states.
In particular, if the excess
electron occupies the state $\left| 0\right\rangle $, no further
electron can be promoted from \ the valence
band into that state, and thus creation of an exciton by a $\sigma ^{+}$%
-polarized laser pulse is inhibited (left part of Fig.
\ref{Pauliblocking}). This effect, referred to as Pauli blocking,
has been experimentally verified \cite{Warbuton97}. On the other
hand, if the excess electron was in $\left| 1\right\rangle $, \
nothing could prevent a second electron from being excited to the state $%
\left| 0\right\rangle $, thereby creating the trion state $\left|
x^{\uparrow }\right\rangle $ (right part of Fig.  \ref{Pauliblocking}) .%
Taking now into account all of the above approximations and
selection rules, we can write the following effective Hamiltonian
for the dynamics of the relevant degrees of freedom in a frame
rotating at the laser frequency $\omega _{L}$:
\begin{eqnarray}
H_{\mathrm{eff}}&=&\hbar \sum_{\nu =a,b}\delta \left|
1\right\rangle _{\nu }\left\langle 1\right| -\Delta (t)\left|
x^{\uparrow }\right\rangle _{\nu }\left\langle x^{\uparrow
}\right| \nonumber\\
&&\mbox{}+\left[ \frac{\Omega (t)}{2}\left| x^{\uparrow
}\right\rangle _{\nu }\left\langle 1\right| +\mathrm{h.c.}\right]
\label{Heff}\\
&&\mbox{}+\Delta E_{ab}\left| x^{\uparrow }\right\rangle
_{a}\left\langle x^{\uparrow }\right| \otimes \left| x^{\uparrow
}\right\rangle _{b}\left\langle x^{\uparrow }\right| \nonumber,
\end{eqnarray}
where for generality we have considered a chirped laser, i.e. one
having a time-dependent detuning $\Delta (t)$ from the $\left|
1\right\rangle \rightarrow \left| x^{\uparrow }\right\rangle $
transition. We have added a global (i.e., independent of the QD
label) splitting $\delta$ between the two logical states, which
can be realized, e.g., via an external static magnetic field. In
the adiabatic scheme we are going to develop, this will have the
effect of suppressing unwanted transitions between $|0\rangle$ and
$|1\rangle$. This few state model will be valid for $\Omega$,
$\Delta$, $\Delta E_{ab}/\hbar$, and the Fourier width of the
pulse $\tau^{-1}$ to be much smaller than the QD level spacing, so
that transitions to excited states are negligible. For $\Delta
(t)=0$, $H_{\mathrm{eff}}$ has the same form as Eq. (\ref{Hphys}),
and is thus suitable for quantum gate operation. A non-zero
$\Delta (t)$ will turn out to be relevant for correcting the
so-called hole-mixing problem, which is outlined in the following
section.

\subsection{Hole mixing}
Indeed, Eq. (\ref{Heff}) does not account
for an important feature of a real QD system, namely the
interaction between the hole subbands, described by the Luttinger
Hamiltonian \cite{Luttinger}. In this more accurate description,
the actual hole eigenstates are no longer the ones described
above. In \ particular, the eigenstate of the (heavy) hole
involved in the dynamics relevant to our study has to include a
correction
from the light-hole state $d_{\nu ,0,+1/2}^{\dagger }\left| \mathrm{vac}%
\right\rangle $. Now, a pulse of $\sigma ^{+}$-polarized light can
promote an electron from the valence-band state corresponding to
such light hole\ into the state $\left| 1\right\rangle $. This
means that the same laser we have included in the Hamiltonian Eq.
(\ref{Heff}) has a certain probability amplitude to excite an
exciton in the QD even if the initial excess electron state was
$\left| 0\right\rangle $, that is, the laser-coupling selection
rules discussed above will be weakly violated in a real QD. This
effect can be included in our simplified three-level model as an
additional coupling between the states $\left| 0\right\rangle $
and $\left| x^{\uparrow }\right\rangle $, induced by the laser
with Rabi frequency $\Omega (t)$ and weighted by the effective
parameter $\varepsilon $ whose typical value is at most 10\%. This
leads to the model Hamiltonian
\begin{eqnarray}
H_{\mathrm{mix}}&=&\hbar \sum_{\nu =a,b}\delta \left|
1\right\rangle _{\nu }\left\langle 1\right| -\Delta (t)\left|
x^{\uparrow }\right\rangle _{\nu }\left\langle x^{\uparrow
}\right| \nonumber\\
&&\mbox{}+\left[ \frac{\Omega (t)}{2}\left( \left| x^{\uparrow
}\right\rangle _{\nu }\left\langle 1\right| +\varepsilon \left|
x^{\uparrow }\right\rangle _{\nu }\left\langle 0\right| \right)
+\mathrm{h.c.}\right]\qquad\label{Hmix}
\\
&&\mbox{}+\Delta E_{ab}\left| x^{\uparrow }\right\rangle
_{a}\left\langle x^{\uparrow }\right| \otimes \left| x^{\uparrow
}\right\rangle _{b}\left\langle x^{\uparrow }\right| ,\nonumber
\end{eqnarray}
which will be the basis for our simulations.

\section{Two-qubit gate implementation}

In this Section we show how the transformation Eq.~(\ref{defGate})
can be realized in practice using quantum dots. We will discuss
the ideal scenario of perfect Pauli blocking as described in
Eq.~(\ref{Heff}), and then introduce imperfections. We will first
lay out a strategy to overcome hole mixing, based on adiabatically
chirped laser pulses. Then, in the next Section, we will show that
the same strategy allows for suppressing the effect of phonon
decoherence.

\subsection{Ideal gate: $\varepsilon=0$}

In the absence of hole mixing, the laser excitation of the trion
is perfectly state-selective. In this ideal case, gate operation
is particularly straightforward.
\subsubsection{Direct Rabi excitation\label{sec:Rabi}}
\begin{figure}[h]
\begin{center}
\epsfig{figure=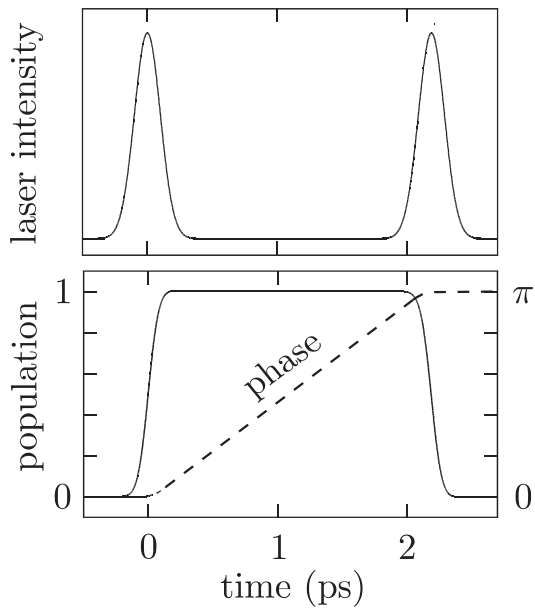,width=7truecm}
\caption{\label{Rabi}Two-qubit gate via direct Rabi excitation:
pulse sequence for exciting and de-exciting the trion state in
each dot (above); trion-trion population and accumulated
interaction phase (below).}
\end{center}
\end{figure}

The simplest quantum gate scheme exploiting the interaction
Eq.~(\ref{Heff}) is based on the following procedure:
\begin{enumerate}
\item selectively excite a trion via a resonant $\pi$ Rabi
rotation; \item wait a sufficient time $\tau\approx\pi\hbar/\Delta
E_{ab}$ for the gate phase $\pi$ to be accumulated; \item
de-excite with a second $\pi$ pulse to return to the logical
subspace.
\end{enumerate}
This is depicted in Fig.~\ref{Rabi}.
Since the trion-trion splitting can give an interaction time scale
of the order of ps, such a gate would be pretty fast -- however,
unfortunately the scheme works only in the idealized model of
perfect Pauli blocking. Therefore we want to develop an
alternative excitation scheme, to be reliable also in the presence
of hole mixing. This can be achieved by employing an adiabatic
technique.

\subsubsection{Adiabatic passage via dressed states\label{sec:dressed2}}
\begin{figure}[h]
\begin{center}
\epsfig{figure=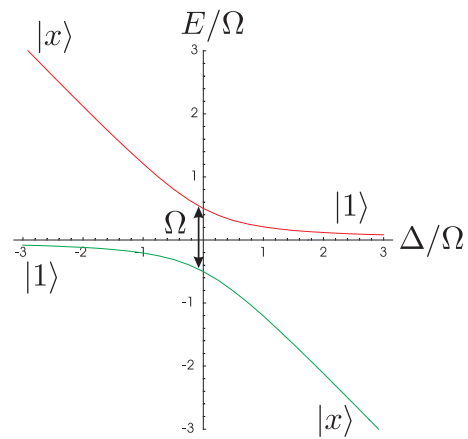,width=6truecm}
\caption{\label{dressed2}Dressed states in a driven two-level
system.}
\end{center}
\end{figure}

We want to design a process which allows to ``switch on'' the
excitonic state for a certain time, and then to return to the
initial ground state with the highest possible probability, by
avoiding at the same time spontaneous emission. We can achieve
this by using an adiabatically chirped laser pulse, i.e. one with
a slowly changing detuning from the excitonic transition. Let us
start by considering our driven two-level system
\begin{equation}
H_{0} =-\Delta \left| x\right\rangle \left\langle x\right| +\frac{\Omega }{%
2}\left| 1\right\rangle \left\langle x\right| +h.c.
\end{equation}
in the strong-coupling regime. Its two eigenstates (the so-called
dressed states)
\begin{eqnarray}
|+\rangle  & = & \sin \frac{\theta }{2}|1\rangle +\cos \frac{\theta }{2}|x\rangle ,\label{dressed+}\\
|-\rangle  & = & \cos \frac{\theta }{2}|1\rangle-\sin \frac{\theta
}{2}|x\rangle ,\label{dressed-}
\end{eqnarray}
where \begin{equation} \tan \theta =-\Omega /\Delta \qquad (0\leq
\theta \leq \pi), \end{equation}
 have the energies
\begin{equation} E_{\pm }=-\frac{\Delta }{2}\pm
\frac{1}{2}\sqrt{\Delta ^{2}+\Omega
^{2}},\label{eq:adiaen}\end{equation}
which are drawn in
Fig.~\ref{dressed2}.
Spontaneous emission can occur only if the system has a non
vanishing probability amplitude of finding itself in its excited
state. On the other hand, for small values of $\Delta/\Omega$ the
dressed ground state contains a significant component of the
interacting state $|x\rangle$. Hence, the following
spontaneous-emission avoiding excitation procedure can be devised:
\begin{enumerate}
\item the system is prepared in the electronic ground state; \item
the laser excitation starts at a large negative value of
$\Delta/\Omega$, where the lower dressed state tends to
$|1\rangle$; \item the laser parameters are slowly changed towards
smaller values of $\Delta/\Omega$, achieving the transformation
\begin{equation} \alpha \left| 0\right\rangle +\beta
\left| 1\right\rangle \rightarrow \alpha \left| 0\right\rangle
+\beta \left( \cos \frac{\theta }{2}\left| 1\right\rangle -\sin
\frac{\theta }{2}\left| x\right\rangle \right)\label{adiabexc};
\end{equation}
\item the system is adiabatically driven back to its initial
state.
\end{enumerate}
If such a chirped laser pulse is applied to two neighboring dots,
they will still acquire a state-dependent trion-trion phase, due
to the admixture from state $|x\rangle$ that is reached starting
from state $|1\rangle$. The main constraint is that we want to
change the Hamiltonian slowly enough such as to remain in the
lower dressed state of the driven system with a high probability.
We will show below that it is even possible to perform the
operation in such a way that no phonon mode is excited during the
gate, thus greatly reducing its sensitivity to temperature.

\subsection{Hole-mixing-tolerant gate}
In the presence of hole mixing, Pauli blocking does not work
perfectly. Therefore a $2\pi$ pulse for the transition
$|1\rangle\to|x^\uparrow\rangle$ will leave behind some excitonic
population, causing decoherence. In this case only an adiabatic
gate operation procedure can ensure that no excitonic population
survive after gate operation. Let us therefore analyze it in more
detail in the case when $\varepsilon\not=0$.

\subsubsection{Hole-mixing-tolerant laser excitation}
\begin{figure}[h]
\begin{center}
\epsfig{figure=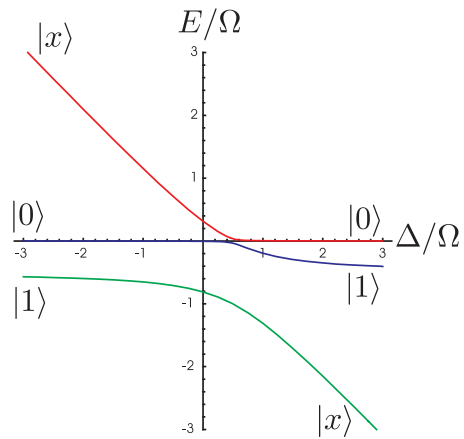,width=6truecm}
\caption{\label{dressed3}Dressed states in a driven three-level
system. A typical hole mixing parameter value of $\varepsilon=0.1$
is assumed.}
\end{center}
\end{figure}

The Hamiltonian in this case is
\begin{equation}
H_{1} =H_{0}+\delta \left| 1\right\rangle \left\langle 1\right| +\frac{%
\varepsilon \Omega }{2}\left| 0\right\rangle \left\langle x\right|
+h.c.
\end{equation}
The level scheme for $H_1$ is drawn in Fig. \ref{dressed3}. It
shows an avoided crossing of the order of $\Omega$ between states
$|1\rangle$ and $|x\rangle$ like in the two-level case, as well as
a much smaller one between $|0\rangle$ and $|x\rangle$. However,
the latter is found at more positive values of the detuning.
Therefore, the same procedure as outlined above will work also in
this case, while the adiabatic excitation Eq. (\ref{adiabexc})
will be still approximately satisfied.

\subsubsection{Two-qubit gate via chirped pulse}
\begin{figure}[h]
\begin{center}
\epsfig{figure=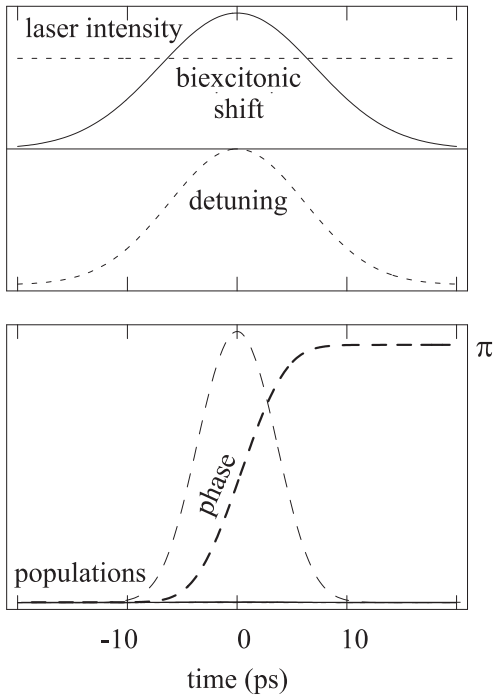,width=6.5truecm}
\caption{\label{adiabgate}Hole-mixing-tolerant two-qubit gate
operation via a chirped laser pulse: pulse shapes (above);
interaction phase and trion-trion populations (below). The dashed
line depicts the population in the $|xx\rangle$ state obtained
starting from $|11\rangle$; the flat solid line the ones obtained
starting from $|00\rangle$, $|01\rangle$ or $|10\rangle$.}
\end{center}
\end{figure}

Let us now have a closer look to what happens when two neighboring
dots undergo the above mentioned pulse sequence. The two-dot
Hamiltonian, including trion-trion interaction, is
\begin{equation} H_{2}
=H_{1}^{a}+H_{1}^{b}+\Delta E_{ab}\left| x\right\rangle
_{a}\left\langle x\right| \otimes \left| x\right\rangle
_{b}\left\langle x\right|.
\end{equation}
\begin{figure}[h]
\begin{center}
\epsfig{figure=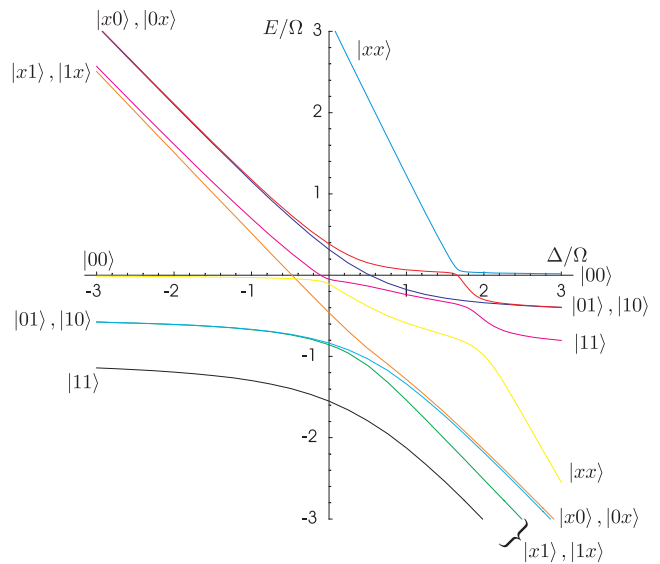,width=8.5truecm}
\caption{\label{dressed9}Level scheme of a system of two coupled
quantum dots including hole mixing.}
\end{center}
\end{figure}
The level scheme is depicted in Fig.~\ref{dressed9}. Although it
is of course significantly more complicated than the single-dot
scheme of Fig.~\ref{dressed3}, the basic feature remains
unchanged: under the same procedure as in the previous section,
with the same laser pulse on both dots, different computational
basis states will acquire different trion-trion amplitudes. This
state selectivity allows one to obtain a nontrivial gate phase
even in the presence of hole mixing, still avoiding spontaneous
emission as discussed in Sec.~\ref{sec:dressed2}. We carried on a
simulation using the following pulse shapes:
\begin{eqnarray}
\Omega(t)&=&\Omega_0e^{-(t/\tau_\Omega)^2},\\
\Delta(t)&=&\Delta_\infty\left[1-e^{-(t/\tau_\Delta)^2}\right].
\end{eqnarray}
Results of the simulation with $\Delta E_{ab}=2$~meV,
$\hbar\Delta_\infty=3$~meV, $\tau_\Omega=10$~ps,
$\tau_\Delta=8.72$~ps, $\delta=0.5$~meV and $\hbar\Omega_0=3$~meV,
are reported in Fig.~\ref{adiabgate}. The gate phase
$\vartheta=\pi$ is obtained in a longer time than in the simple
Rabi-flopping scheme (Sec.~\ref{sec:Rabi}), since the procedure
now has to be adiabatic to avoid excitations to decaying states.
Indeed, the population left in the unwanted excitonic states
remains below $10^{-6}$.

\section{Single-qubit operations\label{sec:singlequbit}}

Contrary to atomic QC implementation schemes, implementing the
single-qubit gate employing the spin state of an electron in a QD
is of greater difficulty than implementing the two-qubit gate
\cite{DiVincenzoNature}. A natural candidate for an optical
implementation of the spin rotation would be employing a
two-photon Raman process involving an intermediate hole state.
However, such a scheme is not possible since the lowest lying
(long lived) hole states have $M_J^h=\pm 3/2$ symmetry, and there
is, however, no strong dipole allowed two-photon coupling
connecting the qubit states. This problem can be overcome by
applying a transverse magnetic field, mixing the spin states and
addressing the Zeeman split states in a frequency selective way,
which however severely limits the Rabi frequencies and thus the
gating time \cite{Imamoglu00}. An alternative is to make a Raman
process via the lowest light hole states $M_J^h=\pm 1/2$ which
however, being excited hole states, suffer from significant
decoherence.

\begin{figure}[h]
\begin{center}
\epsfig{figure=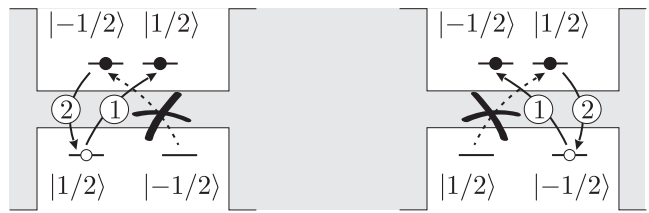,width=8.5truecm}
\caption{\label{singlequbit}Single-qubit operation.}
\end{center}
\end{figure}

Employing II-VI semiconductors grown QDs avoids the above
mentioned decoherence problems, since in these QDs the strain can
shift the energy of the light holes to become the energetically
lowest hole states \cite{light2}. Exploiting such QDs the
single-qubit gate can be performed by the following pulse
sequence: (i) a linearly polarized laser pulse couples the
light-hole subband and the bottom of the conduction band (see
Fig.~\ref{singlequbit}).
This linearly polarized pulse, which can be described as an
equal-weighted superposition of $\sigma_+$ and $\sigma_-$
polarized light, attempts to create a trion-trion state in which
both ground state electronic spin states are occupied but due to
Pauli principle, a $\pi$-pulse of such a light will promote an
electron only into the unoccupied qubit state -- processes (1) in
Fig.~\ref{singlequbit}. (ii) A further $\pi$-pulse of $\sigma^+$
light now recombines the hole state with the original excess
electron Fig.~\ref{singlequbit}. In order to perform this
recombination, i.e., changing the total angular momentum
$\mathbf{j}$ by one unit and keeping its component along the QD\
symmetry axis $\hat{z}$ fixed, the laser pulse is shined in
in-plane direction. This is achieved by employing a wave guide
based scheme \cite{Weiner85}. The Hamiltonian describing the above
process is given by:
\begin{eqnarray}
H_{\rm 1q}&=& \hbar\frac{\Omega_1(t)}2 \left( |{x_l}^\downarrow
\rangle \langle 0| + \varepsilon |{x_l}^\uparrow \rangle \langle
0|\right.\nonumber\\&& \mbox{} +
\left.|{x_l}^\uparrow\rangle\langle 1|+ \varepsilon
|{x_l}^\downarrow\rangle\langle
1|+ {\rm h.c.} \right)\nonumber\\
&&\mbox{}+ \hbar\frac{\Omega_2(t)}2 \left(|{x_l}^\uparrow \rangle
\langle 0| + |{x_l}^\downarrow \rangle \langle 1| + {\rm
h.c.}\right)\nonumber{}\\&& \mbox{} -\Delta(t) \left
(|{x_l}^\downarrow\rangle \langle {x_l}^\downarrow |+
|{x_l}^\uparrow\rangle \langle {x_l}^\uparrow | \right) \
,\nonumber
\end{eqnarray}
where $\left |{x_l}^\uparrow \right\rangle \equiv
\left|{x_l},+1/2\right\rangle, \left |{x_l}^\downarrow
\right\rangle \equiv \left| {x_l},-1/2\right\rangle $ the $l$
index notation is used to indicate that the excitonic states are
defined in terms of light holes, rather than of heavy holes as
before. The electron-electron interaction appearing in the
intermediate state has been absorbed by redefining the detuning.
One should note that the heavy-light hole mixing has an effect
only for the laser pulse directed in the growth direction,
$\hat{z}$, the in-plane directed laser pulse does not induce
transitions for the heavy hole part of the mixed wave function due
to symmetry considerations. By properly adjusting the duration and
the phase of the laser pulses, i.e., adiabatically eliminating the
excitonic states, the following  effective Hamiltonian is
obtained: \begin{equation}H_{\rm 1q}^{eff}= \hbar
\frac{\Omega_{eff}(t)}2 \left| 1\rangle \langle 0\right|+{\rm
h.c.},\end{equation} where $\Omega_{eff}=\sqrt{(\Omega_1 \Omega_2
+ \varepsilon \Omega_1^2)/\Delta}$ is the effective Rabi frequency
for the coherent rotation between the two logical states. Thus
utilizing strain inverted heavy-light hole II-VI semiconductors
grown material, on which the adiabatic two qubit gate procedure
can be adopted, single qubit optical gating can be induced on a
pico-second time scale.

\section{Phonons as a source of decoherence}

\subsection{Interaction of quantum dots with phonons}

The fidelity of the proposed qubit turns out to be quite good, and is
determined by several mechanisms. Since the expected result is
quite small, we can consider different sources of infidelity
separately.

Due to the potentials confining electrons in all spatial
dimensions which lead to a discrete density of states QDs are
often referred to as ``artificial atoms''. The major difference
with respect to atoms is the coupling of the electrons to
underlying lattice degrees of freedom, which lead to relatively
faster decoherence times, on the order of tens of picoseconds
\cite{excitondeph}. In what follows we are developing a simple
model of a QD interacting with a thermal bath of phonons.

In a large system of volume $V$ the displacement field is linear
in creation and annihilation operators of phonons,
$b_{j\mathbf{q}}$ and $b_{j\mathbf{q}}^{\dagger }$ respectively.
Every phonon mode is characterized by its polarization $j$,
momentum $\mathbf{q}$ and frequency $\omega _{j}(\mathbf{q})$. The
number of phonon degrees of freedom is limited by the total number
of atoms forming the lattice. Hence, there is a maximum frequency,
usually defined via Debye temperature $\Theta $, so that $\omega
_{j}(\mathbf{q})\alt \Theta $ (hereafter we use units such that
$\hbar =k_{B}=1$). Normally $\Theta $ is very high (hundreds of
K).

The minimal model describing the interaction of a charged quantum
dot with the phonon field is given by the following Hamiltonian:
\begin{equation}
H_{\rm ph}=H_{\rm QD}-\mathbf{M}_{0}\mathbf{E}(c^{\dagger
}h^{\dagger }+hc)+\sum _{j,\mathbf{q}}\omega
_{j}(\mathbf{q})b_{j\mathbf{q}}^{\dagger
}b_{j\mathbf{q}}+V_{int}.\end{equation} The quantum dot is assumed
to be in a time-dependent laser field $\mathbf{E}$.
$\mathbf{M}_{0}$ is the dipole moment, and the Hamiltonian of an
isolated QD is given by\begin{equation} H_{\rm QD}=\epsilon
^{e}c^{\dagger }c+\epsilon^{h}h^{\dagger }h,\end{equation} where
$c$ and $h$ are the operators annihilating a ground-state electron
and hole, respectively. Their spin dependence is not relevant here
and will not be considered in what follows. As above, the
quantities $\epsilon^{e}$ and $\epsilon^{h}$ are the energies of
the single particle electron and hole states, respectively. The
phonon coupling Hamiltonian is
\begin{equation} V_{int}=\sum
_{j,\mathbf{q}}\left[(g_{j\mathbf{q}}^{e}c^{\dagger
}c-g_{j\mathbf{q}}^{h}h^{\dagger
}h)b_{j\mathbf{q}}+c.c.\right],\end{equation} where
$g_{j\mathbf{q}}^{h}$ and $g_{j\mathbf{q}}^{e}$ are the coupling
constants. The dynamics only couples the states $|g\rangle =|{\rm vac}\rangle
$ and $|e\rangle =c^{\dagger }h^{\dagger }|{\rm vac}\rangle$. The
matrix elements in this basis are: $\langle e|c^{\dagger
}c|e\rangle =\langle e|h^{\dagger }h|e\rangle =1$ and $\langle
e|c^{\dagger }h^{\dagger }|g\rangle =\langle g|ch|e\rangle =1$.
These states are basically the logical and auxiliary states of the
three-level model discussed in Sect.~\ref{sec:threelevel} without
the excess electron, namely,
$|0\rangle=c_{0,-1/2}^\dagger|g\rangle$,
$|1\rangle=c_{0,1/2}^\dagger|g\rangle$, and
$|x,\sigma\rangle=c_{0,\sigma}^\dagger|e\rangle$.

Rewriting the laser field in the form: $E=E(t)\cos (\omega
_{0}t)$, where $\omega _{0}$ is the laser frequency and $E(t)$ is
the slow envelope, in the rotating wave approximation we
obtain:\begin{eqnarray} H_{\rm ph}&=&\Bigl[-\Delta +\sum
_{j\mathbf{q}}\lambda
_{j\mathbf{q}}(b_{j\mathbf{q}}+b_{j\mathbf{q}}^{\dagger
})\Bigr]|e\rangle \langle e|\nonumber\\ &&+\frac{\Omega
}{2}(|e\rangle \langle g|+|g\rangle \langle e|)+\sum
_{j,\mathbf{q}}\omega _{j}(\mathbf{q})b_{j\mathbf{q}}^{\dagger
}b_{j\mathbf{q}},\label{eq:iniHam}\end{eqnarray} where $\Delta $
is the laser frequency detuning, $\Omega
=|\mathbf{M}_{0}\mathbf{E}(t)|$ is the Rabi frequency, and the
phonon coupling strength is given by $\lambda
_{j\mathbf{q}}=g_{j\mathbf{q}}^{e}-g_{j\mathbf{q}}^{h}$. This is
identical, apart from the cavity terms, to the
Imamo\u{g}lu--Wilson-Rae Hamiltonian \cite{WilsonRae} (also used
in Ref.~\cite{kuhn}).

As we shall see, for slow processes considered below the
interaction of a quantum dot with optical phonons is practically
irrelevant. Indeed, optical phonons have a gap: $\omega
_{ opt}(\mathbf{q})\rightarrow \omega _{ opt}\neq 0$
at $q\rightarrow 0$,
which is large and hence optical phonons can always be
adiabatically eliminated from the low frequency dynamics of the
quantum dot. If taken into account, optical phonons contribution
only renormalizes some quantities in the Hamiltonian
(\ref{eq:iniHam}). This is of course not really important for us,
since we consider all the constants in the Hamiltonian as being
phenomenological (essentially taken from experimental data). On
the contrary, longitudinal acoustic phonons (LA phonons) have no
gap: $\omega (\mathbf{q})=uq$ at $q\rightarrow 0$ ($u$ is the
velocity of sound) and can effectively interact with low-frequency
degrees of freedom of the Hamiltonian (\ref{eq:iniHam}). Hereafter
we will only consider LA phonons and omit the index $j$
everywhere.

The Hamiltonian (\ref{eq:iniHam}) formally coincides with the
Hamiltonian of the dissipative spin-boson model. The latter describes the
interaction of a two-level system (spin) with the bath of harmonic
oscillators (bosons). As described in \cite{Leggett}, the
important properties of the interaction of the quantum dot with
the phonons are contained in the integrated quantity
\begin{equation} J(\omega )=\sum _{\mathbf{q}}\lambda
_{\mathbf{q}}^{2}\delta [\omega -\omega
(\mathbf{q})].\label{eq:Jdef}\end{equation}
 One of the most important properties of the spectral function $J$
is its frequency dependence at small $\omega $: $J(\omega )\sim
\omega ^{s}$. The different values of the exponent $s$ distinguish
between the cases of ohmic ($s=1$), sub- ($s<1$) and superohmic
($s>1$) couplings.

The phonons are coupled to the charge distribution in a quantum
dot by means of either deformation or piezoelectric coupling
potentials. The calculation of the spectral function $J$ requires
a specific microscopic model. In the simplest case of a quantum
dot characterized by a harmonic confinement potential the
calculation is pretty easy, even if the QD is placed in external
electric field (see Appendix
\ref{sec:Coupling-constants:-Microscopic}). The results of the
calculation can be summarized as follows: both in the case of
deformation and piezoelectric coupling the spectral function is
superohmic, with $s=3$ and $s=5$, respectively. In both of the
cases the spectral function can be approximately written as
\begin{equation}
J(\omega )\sim \omega ^{s}\exp (-\omega ^{2}/\omega
_{l}^{2})\end{equation} with a cut-off at $\omega _{l}\sim u/l$,
where $l$ is the size of the quantum dot. This frequency is
nothing else but the inverse phonon flight time through the
quantum dot. The electric field (of reasonable intensity) does not
change the exponent $s$ of the spectral function.

\subsection{Adiabatic Hamiltonian: Dressed states}

Instead of considering the {}``bare'' states $|e\rangle $ and
$|g\rangle $ it is convenient to switch to the {}``adiabatic
basis'': let us diagonalize first the quantum dot part of the
Hamiltonian (\ref{eq:iniHam}). The eigenstates (in terms of the
bare states $|1\rangle$ and $|x\rangle$, which in the present
discussion are replaced by $|g\rangle$ and $|e\rangle$,
respectively) and energies are given in Sec.~\ref{sec:dressed2}.

The full interacting Hamiltonian (\ref{eq:iniHam}) can be
rewritten in the new basis and split into two parts: $H_{\rm
ph}=H_{d}+H^{\prime }$, where\begin{eqnarray}
H_{d}&=&\Bigl[E_{+}+\sum _{\bf q}\omega ({\bf q})b_{\bf
q}^{\dagger }b_{\bf q}+\cos ^{2}\frac{\theta }{2}\lambda _{\bf
q}(b_{\bf q}^{\dagger }+b_{\bf q})\Bigr]|+\rangle \langle +|\nonumber\\
&+&\Bigl[E_{-}+\sum _{\bf q}\omega ({\bf q})b_{\bf q}^{\dagger
}b_{\bf q}-\sin ^{2}\frac{\theta }{2}\lambda _{\bf q}(b_{\bf
q}^{\dagger }+b_{\bf q})\Bigr]|-\rangle \langle
-|,\nonumber\\\label{eq:Hdiag}\end{eqnarray} and \begin{equation}
H^{\prime }=-\frac{\sin \theta }{2}\sum _{\bf q}\lambda _{\bf
q}(b_{\bf q}^{\dagger }+b_{\bf q})(|+\rangle \langle -|+|-\rangle
\langle +|),\label{eq:Hnondiag}\end{equation} where
$\tan(\theta)=-\Omega/\Delta$ so that $\theta$ is a time dependent
quantity. The roles of the two Hamiltonians $H_{d}$ and $H^{\prime
}$ are very different and will be considered separately below.

The interaction with the phonons can be understood in two ways.
One of the remarkable non-perturbative simplifications can be used
due to the fact that we are only dealing with the superohmic
coupling case. For a slow process (the Hamiltonian parameters
change on a time scale $\tau \omega _{l}\gg 1$) there is a way to
adiabatically eliminate most of the {}``fast'' phonon degrees of
freedom with $\omega _{*}\alt \omega ({\bf q})\ll \omega _{l}$
without relying on the pertubation expansion in phonon couplings
$\lambda _{k}$. The resulting effective Hamiltonian has the same
form as Eqs. (\ref{eq:Hdiag}-\ref{eq:Hnondiag}), but with the
summation restricted only to the phonon modes with $\omega ({\bf
q})\alt \omega _{*}$ and renormalized values of the Rabi frequency
$\tilde{\Omega }$ and the detuning $\tilde{\Delta }$ (see the
Appendix \ref{sec:Adiabatic-effective-Hamiltonian.}). In what
follows we will use both representations on the same footing and
make no distinction between the bare and the renormalized
quantities whenever it is not relevant.

\subsection{Diagonal and off-diagonal channels; Landau-Zener theory}

Our qubit proposal relies on the fact, that the quantum dot stays
in the same adiabatic ''dressed'' state
(\ref{dressed+}),(\ref{dressed-}) under slow variations of
external parameters ($\Omega $ or $\Delta $). Therefore,
transition between the adiabatic states (\ref{eq:adiaen}) are a
source of infidelity. Realistically every gate operation is
performed with a finite speed and thus undesired transitions
between the dressed states are always possible. In the absence of
phonons the transition probability is given by the Landau-Zener
theory. In its simplest version, i.e. for a case of linear
detuning sweep $\Delta =\dot{\Delta }t$ around the resonance value
$\Delta =0$, the measure of infidelity is given by the probability
\cite{LL} \begin{equation} P_{\pm }=\exp (-\pi \Omega
^{2}/4\dot{\Delta }).\label{eq:interLZ}\end{equation} By
requesting this quantity to be small we establish our adiabatic
condition\begin{equation} \eta =\frac{\Omega ^{2}}{\dot{\Delta
}}\gg 1.\label{eq:adiabcrit}\end{equation} The condition has a
simple physical meaning: the resonance is observed approximately
when we have $\Delta \sim \Omega $, so $\tau \sim \Omega
/\dot{\Delta }$ is nothing else but the characteristic time of the
detuning sweep. Then, the adiabatic condition naturally implies
that we have $\Omega \tau \gg 1$.

The effects of the interaction with phonons on this decoherence
channel do not change this result much. As discussed in the
Appendix \ref{sec:Adiabatic-effective-Hamiltonian.}, both in the
case of perturbation theory and in the adiabatic approximation the
phonon interaction only renormalizes the Rabi frequency [see Eq.
(\ref{eq:omegaren})]. Therefore, the same expression
(\ref{eq:interLZ}), but with the renormalized value of
$\tilde{\Omega }$ instead of $\Omega$, holds even if the phonon
coupling is strong.

Now let us consider the interaction with the phonons in some more
detail. The phonon-assisted transitions between the dressed states
$|\pm \rangle $ are possible and are described by the Hamiltonian
(\ref{eq:Hnondiag}). Since most of the transitions occurs close to
the resonance $\Delta =0$, when the characteristic energy
difference between the adiabatic levels is $\sim \Omega $, and the
inter-level transition probability can be estimated using the
Hamiltonian $H^{\prime }$ and the Fermi Golden Rule
\begin{equation} P_{\pm }\sim \int _{\sim \Omega }^{\infty
}|T_{\pm }(\omega )|^{2}J(\omega -\Omega )d\omega
,\label{eq:transestim}\end{equation} where \begin{equation} T_{\pm
}(\omega )=\frac{1}{2}\int dt\exp (i\omega t)\sin [\theta
(t)]\end{equation} is the Fourier component of the ``coupling
potential'' in Eq. (\ref{eq:Hnondiag}). The latter can be easily
estimated in the adiabatic limit by using the linear approximation
for the detuning close to the resonance point (as
above):\begin{equation} T_{\pm }(\omega )\sim \frac{\exp (-\omega
/\omega _{m})}{(\omega \omega _{m})^{1/2}}\end{equation} where we
have $\omega _{m}=\tau ^{-1}=\dot{\Delta }/\Omega $, and $\omega
\gg \omega _{m}$. The frequency $\omega _{m}$ is the
high-frequency cutoff imposed by the speed of the frequency
detuning sweep. Substituting the above expression into Eq.
(\ref{eq:transestim}) and integrating over $\omega $, we find the
following estimation for the interstate transition
probability:\begin{equation} P_{\pm }\sim \frac{J(\omega
_{m})}{\Omega }\exp (-\alpha \Omega ^{2}/\dot{\Delta
}),\label{eq:interphon}\end{equation} where $\alpha \sim 1$ is a
numerical factor. This quantity characterizes the probability of
unwanted processes and hence is a measure of infidelity. The
result is exponentially small if we have $\omega _{m}\ll \Omega $,
which is nothing else but our adiabaticity condition
(\ref{eq:adiabcrit}).

At finite temperatures there is an additional mechanism for
decoherence: a quantum dot can interact with the phonon field and
absorb a thermal phonon. Accordingly, the quantum dot acquires
energy and is transformed  into the excited dressed state.
Nevertheless, the process can be easily suppressed by operating in
the regime of small temperatures $T\ll \Omega $. In this case the
transition probability is characterized by an additional small factor
$\exp (-\Omega /T)\ll 1$ and can be disregarded.

Both probabilities (\ref{eq:interLZ}) and (\ref{eq:interphon})
have similar structure and are exponentially small for adiabatic
processes (\ref{eq:adiabcrit}). In later sections we will find
that the ``diagonal'' terms $H_{d}$ in the quantum dot
Hamiltonian, though  not changing the adiabatic states of the
quantum dot, lead nevertheless to excitations of acoustical
phonons and thus to a certain infidelity. The results do not
contain exponentially small factors and hence the off-diagonal
terms in the Hamiltonian considered here can be neglected in our
further fidelity calculations.

\subsection{General expression for the fidelity in the presence of phonons}

The major source of infidelity is the excitation of acoustical
phonons without change of the quantum dot state,  i.e., pure dephasing.
To develop a
formal approach to the fidelity calculation, we consider the
evolution of a quantum dot coupled to a heat bath of phonons.
Assume that at $t=-\infty $ the system starts from the state which
is a direct product of pure quantum dot state and a thermal state
of phonon field at a temperature $T$. As discussed in the previous
section, the dynamics of the quantum dot can be well described (up
to a few exponentially small amplitudes) by the diagonal
Hamiltonian (\ref{eq:Hdiag}). Using this simplification, we can
rewrite the density matrix of the quantum dot subsystem at
$t=\infty $ as \begin{equation} \rho _{\alpha \beta }=\langle U_{\alpha
}^{\dagger }\overleftarrow{T}\overrightarrow{T}U_{\beta }\rangle
|\alpha \rangle \langle \beta |,\label{eq:dmoper}\end{equation} where
$|\alpha \rangle $ is the quantum dot state, $\langle ...\rangle $ is
the average over the initial phonon state, $\overrightarrow{T}$ is
the time ordering sign, and the (diagonal in the dressed state
basis) evolution operators are given by\begin{equation}
\overrightarrow{T}U_{\alpha }=\overrightarrow{T}\exp \Bigl\{-i\int
dt\Bigl[f_{\alpha }(t)\sum _{k}\lambda _{k}(b_{k}+b_{k}^{\dagger
})+E_{\alpha }\Bigr]\Bigr\},\end{equation} with $f_{+}=\cos
^{2}(\theta(t) /2)$ and $f_{-}=-\sin ^{2}(\theta(t) /2)$. The time
ordering is not convenient and can be removed by transforming the
evolution operator into\begin{equation} \overrightarrow{T}U_{\alpha
}=\exp \Bigl\{-i\int _{-\infty }^{\infty }dt\Bigl[f_{\alpha }(t)\sum
_{k}\lambda _{k}(b_{k}+b_{k}^{\dagger })+E_{\alpha }\Bigr]+i\phi
_{\alpha }\Bigr\},\end{equation} where
\begin{equation} \phi _{\alpha }=\int _{-\infty }^{\infty }dt\int
_{\infty }^{t}dt^{\prime }f_{\alpha }(t)f_{\alpha }(t^{\prime })\sum
_{k}\lambda _{k}^{2}\sin \omega _{k}(t-t^{\prime })\end{equation}
 is the phase originating from the non-commutativity of $b_{k}$ operators
at different times. Combining these results together we obtain the
following expression for the fidelity matrix $T_{\alpha \beta }=\langle
U_{\alpha }^{\dagger }\overleftarrow{T}\overrightarrow{T}U_{\beta
}\rangle $:\begin{equation} T_{\alpha \beta }=\Bigl\langle \exp
\Bigl\{-i\int _{-\infty }^{\infty }dta_{\alpha \beta }(t)\sum
_{k}\lambda _{k}(b_{k}+b_{k}^{\dagger })+i\phi _{\alpha \beta
}\Bigr\}\Bigr\rangle ,\label{eq:fidmatrix}\end{equation} where
$a_{\alpha \beta }=f_{\alpha }-f_{\beta }$, and \begin{eqnarray} \phi _{\alpha
\beta }&=&\phi _{\beta }-\phi _{\alpha }+\int _{-\infty }^{\infty
}dt\left[E_{\beta }(t)-E_{\alpha }(t)\right]\label{eq:phononphase}\\
&&+\iint dtdt^{\prime }f_{\alpha }(t)f_{\beta }(t^{\prime })\sum
_{k}\lambda _{k}^{2}\sin \omega _{k}(t-t^{\prime
}).\nonumber\end{eqnarray}

The fidelity matrix $T$ can be used to rewrite Eq.
(\ref{eq:dmoper}) in a more convenient form. Consider an arbitrary
pure state of a quantum dot $\Psi =\sum _{\alpha }c_{\alpha
}|\alpha \rangle $, which evolves into the density
matrix\begin{equation} \rho _{\alpha \beta }=c_{\alpha
}^{*}c_{\beta }T_{\alpha \beta }(\lambda
_{k}).\label{eq:densmatfin}\end{equation} Without the interaction
with phonons the evolution is characterized by the $T$-matrix with
all $\lambda _{k}$ set to zero. Therefore, one can characterize
the phonon interaction by the degree of infidelity, defined
as\begin{equation} f\equiv1-F=\max _{\{c_{\alpha }\},\sum _{\alpha
}|c_{\alpha }|^{2}=1}\sum_{\alpha\beta}|c_{\alpha }^{*}c_{\beta
}[T_{\alpha \beta }(\lambda _{k})-T_{\alpha \beta
}(0)]|.\label{eq:infidelitydef}\end{equation} This is a standard
problem of linear optimization, whose solution can be done in the
general form: the infidelity is given by the largest eigenvalue of
the matrix $T(\lambda )-T(0)$.

In what follows we will estimate the value of the infidelity $f$
for single qubit operations and for the quantum gate realization
proposed above.

\subsection{Fidelity of Rabi rotation}

We consider first the simplest case and calculate the fidelity of
single qubit operations, such as a reversible internal rotation.
To be specific, we calculate the fidelity of an adiabatic sweep of
the detuning $\Delta $ around its resonant value $\Delta =0$,
while keeping the Rabi frequency $\Omega $ constant. Using our
fidelity definition from Eq. (\ref{eq:infidelitydef}), we find
that we have $f=1-\exp (-\Gamma )$, with \begin{equation} \Gamma
=\frac{1}{2}\int d\omega J(\omega )|a(\omega )|^{2}[1+2N(\omega
)],\label{eq:fid1QD}\end{equation} where $N(\omega )=[\exp (\omega
/T)-1]^{-1}$ is the phonon occupation number, and \begin{equation}
a(\omega )=\int _{-\infty }^{\infty }dt\exp (-i\omega t)\cos
(\theta ).\label{eq:Fcomp}\end{equation} In order to analyze the
QD dynamics and compare the results with the discussion of the
off-diagonal processes we use the same sort of linear
approximation for the time-dependent detuning $\Delta =\dot{\Delta
}t$ close to the resonance. A simple calculation gives
\begin{equation} \Gamma \approx \frac{1}{2}\int \frac{\Omega
^{2}J(\omega )}{\dot{\Delta }^{2}}K_{1}^{2}\left(\frac{\omega
}{\omega _{m}}\right)[1+2N(\omega )]d\omega
,\label{eq:gammaest}\end{equation}
 where we have $\omega _{m}=\dot{\Delta }/\Omega \ll \Omega $ (see the discussion
above), and $K_{1}(x)$ is the Bessel function of the second kind.
The main contribution to the integral originates from the range of
frequencies $\omega \alt \omega _{m}$ (assuming, of course,
$\omega _{m}\ll \omega _{l}$). Therefore, at small temperatures
$T\ll \omega _{m}$ we can neglect the thermal occupation of the
phonon states and find that \begin{equation} \Gamma \sim
\frac{J(\omega _{m})}{\omega _{m}}.\label{eq:gamma0}\end{equation}
In the opposite limit, i.e. when $T\gg \omega _{m}$, the phonon
numbers can be approximated as $N(\omega )\approx T/\omega $ and
the integration yields\begin{equation} \Gamma \sim \frac{J(\omega
_{m})}{\omega _{m}}\frac{T}{\omega
_{m}}.\label{eq:gammainf}\end{equation}
 If the coupling with the phonons is weak, i.e. $\Gamma \ll 1$, then
the infidelity coincides with $\Gamma $ and is only a power law
small (compare Eqs. (\ref{eq:gamma0}) and (\ref{eq:gammainf}) with
the exponentially small results (\ref{eq:interLZ}) and
(\ref{eq:interphon}) of our off-diagonal Hamiltonians discussion).
At zero temperature the conditions $\Gamma \ll 1$ and the
perturbation theory expansion parameter (\ref{eq:expanparam}) are
the same.

Eqs. (\ref{eq:gamma0}) and (\ref{eq:gammainf}) are obtained
assuming that $\omega _{m}\ll \omega _{l}$. In this case the
results are not confined to the perturbation theory limit (see the
Appendix \ref{sec:Adiabatic-effective-Hamiltonian.} for more
details about the adiabatic elimination of high frequency
phonons). In the other limiting case the infidelity is given by
the same Eqs. (\ref{eq:gamma0}) and (\ref{eq:gammainf}) but after
the substitution $\omega _{m}\rightarrow \omega _{l}$. Of course,
the condition $\omega _{m}\agt \omega _{l}$ breaks the adiabatic
separation of the slow and fast phonon degrees of freedom.
Therefore, the results for the fidelity in this regime can only be
valid if the infidelity $\Gamma $ is small.

Eq. (\ref{eq:gammaest}) is derived in such a way, that its
validity is not confined solely to the analysis of acoustical
phonons. In fact it also allows one to understand how the
contribution of the higher frequency degrees of freedom (such as
optical phonons) can be ruled out. Indeed, optical phonons are
characterized by the minimum frequency $\omega _{0}$ (the optical
gap), so that $J(\omega )=0$ for all $\omega <\omega _{0}$.
Substituting this definition into Eq. (\ref{eq:gammaest}) and
integrating in the adiabatic limit $\omega _{m}\ll \omega _{0}$ we
find\begin{equation} \Gamma \sim \frac{J_{0}}{\omega _{0}}\exp
(-\frac{\omega _{0}}{\omega _{m}}),\end{equation} where
$J_{0}=J(\omega _{0})$. This is once again an exponentially small
result (compare with Eqs. (\ref{eq:gamma0}) and
(\ref{eq:gammainf})) with a clear physical meaning: a slow process
occurring on a time scale $\omega _{m}^{-1}$ can not excite high
frequency lattice vibrations if $\omega _{0}\gg \omega _{m}$.

\subsection{Rabi oscillations}

Another revealing example of phonon interaction effects is the
damping of Rabi oscillations. Consider the case of exact resonance
($\Delta =0$) and a quantum dot starting at $t=0$ in the state
$|g\rangle$. In the dressed state picture this corresponds to
\begin{equation} |g\rangle =\frac{1}{\sqrt{2}}(|+\rangle
+|-\rangle ).\end{equation} As  time progresses, the state
changes and the probability to find the quantum dot in the state
$|g\rangle $ can be found using the density matrix from Eq.
(\ref{eq:densmatfin})\begin{equation}
P_{g}=\frac{1}{4}[T_{++}+T_{--}+2\Re T_{+-}].\end{equation} In our
diagonal approximation, at $t\rightarrow \infty $, this is
equivalent to\begin{equation} P_{g}=\frac{1}{2}[1+\cos (\Omega
t+\tilde{\phi }_{+-})],\end{equation} with $\tilde{\phi }$ given
by the first two terms in Eq. (\ref{eq:phononphase}). This means
that the diagonal Hamiltonian (\ref{eq:Hdiag}) describes undamped
Rabi oscillations at the frequency $\Omega $. Within the adiabatic
approximation $\Omega \ll \omega _{l}$ one can integrate out the
high frequency phonons (see Appendix
\ref{sec:Adiabatic-effective-Hamiltonian.}) and observe that in
the first approximation the effects of phonon interaction show up
in the renormalization of the Rabi oscillation frequency
$\Omega \rightarrow \tilde{\Omega }$, as given by Eqs.
(\ref{eq:omegaren}), (\ref{eq:omegarenT}).

The gradual damping of the Rabi oscillations originates from the
off-diagonal Hamiltonian $H^{\prime }$ (\ref{eq:Hnondiag}). In
contrast to our previous discussion of the internal qubit
rotation, in this case there is a finite probability to find the
quantum dot in its excited state $|+\rangle $. This means that now
the processes leading to emission of phonons become possible. The
transition rate $\Gamma $ can be calculated using the Fermi Golden
Rule\begin{equation} \Gamma \sim J(\Omega ),\end{equation} so that
\begin{equation} P_{g}\approx \frac{1}{2}[1+\cos (\Omega
t+\tilde{\phi }_{+-})\exp (-\Gamma t)].\end{equation} Since
realistically it is $\Omega \ll \omega _{l}$, we have the ratio
$J(\Omega)/ \Omega \ll 1$ and thus the quantum-dot oscillations
are only weakly damped.

Altogether this lets us conclude that in the presence of phonons a
quantum dot in an external laser field undergoes weakly damped Rabi
oscillations, characterized by the renormalized frequency and the
damping rate determined by the spectral function $J(\Omega )$.

\subsection{Quantum Gate fidelity}

The ultimate goal of our calculations is the fidelity of a quantum
gate. The Hamiltonian of a couple of interacting QDs can be
represented as follows\begin{eqnarray} H_{\rm
ph}^{(2)}&=&\Bigl[-\Delta +\sum _{\mathbf{q},\nu }\lambda
_{\mathbf{q}}\big(b_{\mathbf{q}}e^{i\mathbf{q}\mathbf{x}_{\nu
}}+b_{\mathbf{q}}^{\dagger }e^{-i\mathbf{q}\mathbf{x}_{\nu
}}\big)\Bigr]|e\rangle _{\nu }\langle e|\nonumber\\
&&+\frac{\Omega }{2}\sum _{\nu }(|e\rangle _{\nu }\langle
g|+|g\rangle _{\nu }\langle e|)+\sum _{\mathbf{q}}\omega
(\mathbf{q})b_{\mathbf{q}}^{\dagger
}b_{\mathbf{q}}\nonumber\\
&&+\Delta E_{ab}|e\rangle _{a}\langle e|\otimes|e\rangle
_{b}\langle e|,\end{eqnarray} where the index $\nu =a,b$ labels
the quantum dots, $x_{\nu }=\pm d/2$ is the position of the dots,
and the last term represents the trion-trion shift $\Delta
E_{ab}$.

\begin{figure}
\epsfig{figure=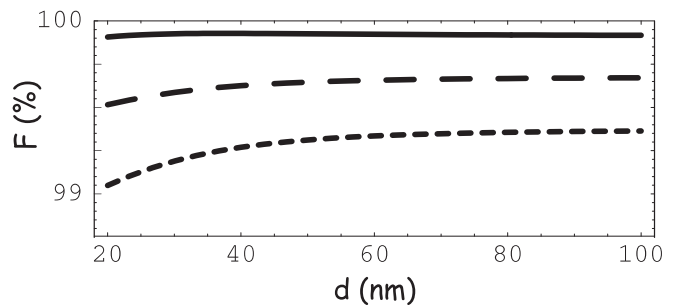,width=\columnwidth} \caption{ The
fidelity of the quantum gate as a function of the inter-quantum
dot separation $d$, for different values of the temperature $T$:
0~K (solid line), 5~K (long-dashed line), 15~K (dashed
line).\label{fig:fidelity-a-T}}
\end{figure}

The trion-trion shift is a crucial element of our quantum gate
proposal. Its presence introduces the conditional dynamics. The
timing of the adiabatic process should be designed in such a way
that the trion-trion shift induces the required phase shift of the
state $|e\rangle _{a}|e\rangle _{b}$.

The whole analysis of a single quantum dot operation can be
easily generalized to the quantum gate case. As discussed above,
we first transform into the dressed state basis and then select
only ``diagonal'' terms from the interaction with the phonons.
Then one can calculate the fidelity matrix $T$ (now $4\times 4$)
and find the fidelity from Eq. (\ref{eq:infidelitydef}). This
program can be completed numerically.

Let us consider first the two simple limiting cases. In the
simplest case of a small trion-trion shift $\Delta E_{ab}\ll
\Omega $, the two quantum dots can be considered separately and
the trion-trion shift can be accounted for as a perturbation.
Then, following the steps of our single quantum dot calculation,
we find that the infidelity is $f=1-\exp (-\Gamma )$,
with\begin{equation} \Gamma =\int J(\omega )d\omega \cos
^{2}(\omega d/u)[1+2N(\omega )]|a(\omega
)|^{2},\label{eq:fid2QD}\end{equation}
 where $a(\omega )$ is again given by Eq. (\ref{eq:Fcomp}). As expected,
in this limit the result is practically the same as we had for a
single qubit operation -- see Eq. (\ref{eq:fid1QD}).

The dependence of the gate fidelity (\ref{eq:fid2QD}) on the
quantum dots separation $d$ is very weak (the effective value of
the $\cos $-function under the integral sign is anything between
$1/2$ and $1$ for large and small values of $d$, respectively).
This means that the obtained result is practically insensitive to
the assumptions regarding the degree of coherence between the
phonon modes around each of the dots. Indeed, Eq.
(\ref{eq:fid2QD}) implies that both of the quantum dots interact
with the same phonon bath. This corresponds to a case of having
the dots interacting with the same set of bulk modes. Another
possibility is that the inter-quantum dot separation is smaller
than the phonon mean free path. In the case of separate phonon
baths (which means the quantum dots electronic degrees of freedom
interact with independent phonon modes, or the separation between the
dots exceeds the mean free path of phonons), the quantum dots
become completely separate and the infidelity is given by its
single-qubit expression (\ref{eq:fid1QD}). Both expressions are
indistinguishable within a factor $\sim 1$. The weak dependence of
the fidelity on the interdot separation at not-too-high $T$ can be
seen on the Fig. \ref{fig:fidelity-a-T}, obtained by our exact
numerical calculation.

\begin{table}[h]
\begin{tabular}{|c|c|c|c|}
\hline $\Omega$ (meV)& $f(T=0.1$ K)& $f(T=4$ K)&
$f(T=20$ K)\\
\hline \hline 0.3& $0$& $2.2\div 1.9\times 10^{-3}$&
$5.9\div4.5\times 10^{-3}$\\
\hline 0.06& $0$& $9.7\div4.6\times 10^{-5}$&
$4.8\div2.3\times 10^{-4}$\\
\hline 0.02& $0$& $7.1\div5.1\times 10^{-6}$&
$5.3\div2.5\times 10^{-5}$\\
\hline
\end{tabular}
\caption{ The infidelity $f(T)$ as a function of the temperature
$T$ for different values of the Rabi frequency $\Omega$. The dot
size is $20$ nm. The trion-trion shift $\Delta E_{ab}$ varies from
$0.1$ meV to $3$ meV.\label{table:a20}}
\end{table}
\begin{table}[h]
\begin{tabular}{|c|c|c|c|}
\hline $\Omega$ (meV) & $f(T=0.1$ K)& $f(T=4$ K)&
$f(T=20$ K)\\
\hline \hline 0.3& $0$& $5.6\div4.1\times 10^{-3}$&
$0.034\div0.020$\\
\hline 0.06& $0$& $5.0\div4.6\times 10^{-5}$&
$2.5\div2.3\times 10^{-3}$\\
\hline 0.02& $0$& $1.5\div1.4\times 10^{-5}$&
$7.2\div6.9\times 10^{-5}$\\
\hline
\end{tabular}
\caption{ The infidelity $f(T)$ as a function of the temperature
$T$ for different values of the Rabi frequency $\Omega$. The dot
size is $15$ nm. The trion-trion shift $\Delta E_{ab}$ varies from
$0.1$ meV to $3$ meV.\label{table:a15}}
\end{table}
\begin{table}[h]
\begin{tabular}{|c|c|c|c|}
\hline $\Omega$ (meV) & $f(T=0.1$ K)& $f(T=4$ K)&
$f(T=20$ K)\\
\hline \hline 0.3& $0$& $1.4\div0.92\times 10^{-2}$&
$0.15\div0.1$\\
\hline 0.06& $0$& $1.1\times 10^{-4}$&
$1.2\times 10^{-3}$\\
\hline 0.02& N/A& N/A&
N/A\\
\hline
\end{tabular}
\caption { The infidelity $f(T)$ as a function of the temperature
$T$ for different values of the Rabi frequency $\Omega$. The dot
size is $10$ nm. The trion-trion shift $\Delta E_{ab}$ varies from
$0.1$ meV to $3$ meV.\label{table:a10}}
\end{table}

The calculation in the other limiting case $\Delta E_{ab}\gg
\Omega $ is very similar. In this case the quantum gate has two
avoided crossings instead of one and their contributions add
independently. Since the width of
the resonance is in both cases $\Omega $, the contributions of
independent resonances add separately and are again of the same
order of magnitude as Eq. (\ref{eq:gamma0}) and
(\ref{eq:gammainf}). The dependence of the fidelity on the
interdot separation is very weak again.

To quantify the discussion above we performed the exact numerical
calculation of the fidelity matrix (\ref{eq:fidmatrix}) using a
certain shape of the detuning sweep $\Delta (t)$. The fidelity as
a function of the interdot separation and of the temperature for a
specific value of the trion-trion shift is plotted in
Fig.(\ref{fig:fidelity-a-T}). The figure nicely shows both
temperature regimes (\ref{eq:gamma0}) and (\ref{eq:gammainf}), as
well as the weak dependence of the result on the separation
between the dots. We also performed a few calculations for quantum
dots of different sizes (effectively varying the cutoff parameter
$\omega _{l}$). The results of the calculations are summarized in
the Tables (\ref{table:a20})-(\ref{table:a10}). The interparticle
separation is 5 nm, the pulse duration is 1 ps.

\subsection{Discussion}

In this section we performed a systematic study of various
decoherence mechanism associated with the interaction of the
quantum dots with phonons. The results of our study can be
summarized as follows.

The details of the interaction with phonons can be
{}``compressed'' into the spectral function $J(\omega )$. It is
characterized by the strength of the coupling, the frequency
dependence at small $\omega $ and the value of the high
frequencies cutoff $\omega _{l}\sim u/l$, where $u$ is the
velocity of sound and $l$ is the size of the quantum dot.

The infidelity $f$ turns out to be quite good for all the
realistic situations we considered. This means that the
perturbation theory is well applicable and, in the limit of small
temperatures, the infidelity $f=1-\exp\{-\Gamma\} $ is given by
\begin{equation} \Gamma \sim \frac{J(\omega _{c})}{\omega
_{c}},\label{eq:Gfin0}\end{equation} where $\omega _{c}=\min
(\omega _{l},\omega _{m})$ and $\omega _{m}\sim \dot{\Delta
}/\Omega $ is the inverse characteristic time of the gate
operation. In the higher temperature limit ($T\gg \omega _{c}$)
the infidelity scales linearly with the
temperature\begin{equation} \Gamma \sim \frac{J(\omega
_{c})}{\omega _{c}}\frac{T}{\omega
_{c}}.\label{eq:GfinT}\end{equation} These expressions are the
central results of the section. They can be applied to estimate
the fidelities of both single qubit operations and of the quantum
gates (in the latter case there is also a weak dependence on the
separation between the quantum dots). The accuracy of the simple
estimations along the lines of Eqs. (\ref{eq:Gfin0}) and
(\ref{eq:GfinT}) was checked by exact numerical calculations of
the fidelity in a wide parameter range. The contributions of the
piezoelectric coupling are numerically smaller by 2-3 orders of
magnitude in all our calculations.

\section{State read-out by quantum jumps}
A necessary requirement for a quantum information processing
implementation scheme is the ability to perform an accurate
measurement of a single qubit. Implementation of a highly
efficient solid-state measurement scheme designed to measure the
spin or charge of single electron is a highly difficult task
\cite{Milburn01}.

Monitoring the fluorescence from a single quantum dot (QD) has
been suggested as a mean to measure single scattering events
within QDs \cite{Hohenester}, as well as a means for final read
out of the spin state for the purpose of quantum computation
\cite{Imamoglu00}. Recently, it has been verified experimentally
that the spin state of an electron residing in a QD can be read
using circular pumped polarized light \cite{Cortez}. In this
section we describe how it is possible to devise an optical read
out scheme based on the idea of the Pauli-blocking in QDs even in
the presence of heavy/light hole mixing. We describe two
situations: an ideal case with no heavy/light hole mixing and the
realistic case which includes mixing. In the ideal case the time
of measurement, i.e. the time in which one can still extract the
information regarding the spin state of the confined electron, is
limited only by the spin decoherence time whereas in the case of
mixing the measurement time is also limited by the typical time
for a spin flip induced by the excitation process.

\begin{figure}[h]
\begin{center}
\epsfig{figure=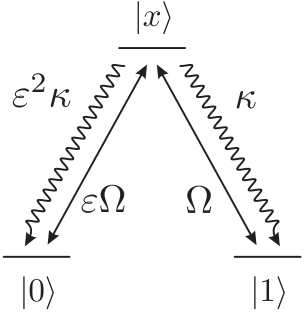,width=4truecm}
\caption{\label{qjumps}The lambda configuration one has to
consider to include hole mixing.}
\end{center}
\end{figure}

The system we have in mind is described by Fig.~\ref{qjumps}. It
is the single-QD counterpart of Eq.~(\ref{Hmix}), taking into
account also the decay rates from the excited level, which we call
$|x\rangle$ here for simplicity. We ignore the detunings $\delta$
and $\Delta$, which are not relevant here, as well as the decay
rate $\kappa_{01}$ from state $|0\rangle$ to $|1 \rangle$, since
the typical time scale for it \cite{excitonspindeph} is much
larger than the typical time scale in which the spin directional
information is lost due to the laser mediated spin flip which is
now the bottleneck process limiting the measurement time.

\subsection{Ideal case, i.e. no mixing} Let us first consider
the case when $\varepsilon=0$. Shining a $\sigma^{+}$ pulse on the
QD we obtain due to the Pauli blocking effect in QDs the usual
two-level situation: no fluorescence from initial state
$|0\rangle$, full fluorescence from state $|1 \rangle$. The final
state measurement i.e. measuring the spin state of the excess
electron in the QD is obtained by the quantum jump technique (e.g.
Ref. \cite{Imamoglu00}): when the original state of the spin in a
QD is $|1 \rangle$ a fluorescence pattern is obtained, whereas the
state $|0 \rangle$ is completely decoupled from the laser field
since exciton creation is blocked by the Pauli principle.

The typical time scale which limits the process is the spin
coherence time in the QD which is of the order of microseconds,
i.e. in a time of that order of a microsecond the spin of the
electron in the QD will flip from the $|1\rangle$ fluorescing
state to the $|0\rangle$ dark state and the fluorescence pattern
will be terminated. The average number of photons emitted before
the spin typically flips its state is given by the ratio of the
spin coherence time to the typical rate for spontaneous emission,
which is of the order of a few nano-seconds. Therefore, typically
one should obtain on the order of $10^3$ photons before the
original spin information is destroyed.

\subsection{Case with mixing} A realistic QD will exhibit mixing
of the heavy and light hole states. This invalidates the
assumption of perfect Pauli blocking with $\sigma^+$ light and can
be viewed as a rotation by an angle $-\varepsilon$ in the $\{|0
\rangle, |1 \rangle\}$ space. The mixing parameter $\varepsilon$
will typically be of the order of the lattice constant, $a$, to a
typical length scale defining the QD in our case $a/L\approx 0.1$
where $L$ is the size of the dot in the growth direction.

\begin{figure}[h]
\begin{center}
\epsfig{figure=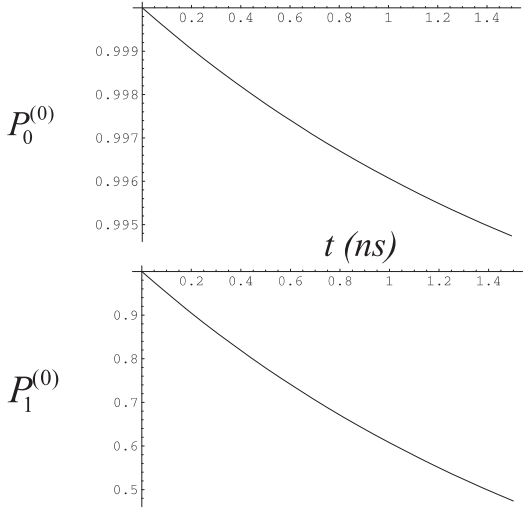,width=7truecm}
\caption{\label{trace}Probability that at time $t$ the first
photon has not yet been emitted, starting from state $|0\rangle$
(above) or $|1\rangle$ (below) at time $t=0$. Parameters are
quoted in the text.}
\end{center}
\end{figure}

Introducing mixing requires one to treat the full three-level
lambda configuration shown in Fig. \ref{qjumps}. As opposed to the
usual atomic lambda configuration \cite{Zoller87}, here one can
not distinguish between the $|0 \rangle \langle x|$ and $|1
\rangle \langle x|$ transitions. These two transitions are
mediated through the same photon. The dissipative evolution of the
density matrix $\tilde\rho(t)$ is given by \cite{Quantumnoise}
\begin{eqnarray}
\dot{\tilde\rho}_{00}&=&i\frac{\Omega}2\varepsilon\left(\tilde\rho_{x0}-\tilde\rho_{0x}\right)\nonumber\\
\dot{\tilde\rho}_{11}&=&i\frac{\Omega}2\left(\tilde\rho_{x1}-\tilde\rho_{1x}\right)\nonumber\\
\dot{\tilde\rho}_{xx}&=&i\frac{\Omega}2\left[\tilde\rho_{1x}-\tilde\rho_{x1}+\varepsilon\left(\tilde\rho_{0x}-\tilde\rho_{x0}\right)\right]-(1+\varepsilon^2)\kappa\tilde\rho_{xx}\nonumber\\
\dot{\tilde\rho}_{01}&=&i\frac{\Omega}2\left(\varepsilon\tilde\rho_{x1}-\tilde\rho_{0x}\right)\label{eqrho}\\
\dot{\tilde\rho}_{0x}&=&i\frac{\Omega}2\left[\varepsilon\left(\tilde\rho_{xx}-\tilde\rho_{00}\right)-\tilde\rho_{01}\right]-(1+\varepsilon^2)\frac\kappa
2\tilde\rho_{0x}\nonumber\\
\dot{\tilde\rho}_{1x}&=&i\frac{\Omega}2\left(\tilde\rho_{xx}-\tilde\rho_{11}-\varepsilon\tilde\rho_{10}\right)-(1+\varepsilon^2)\frac\kappa
2\tilde\rho_{1x}\nonumber
\end{eqnarray}
The probability that at time $t$ no photon has been emitted,
starting from state $\alpha$ at time $t_0$, is
\begin{equation}
P^{(0)}_\alpha(t-t_0)={\rm tr}\left[\tilde\rho(\alpha,t)\right],
\end{equation}
where at the initial time $t_0$ we take
$\tilde\rho(\alpha,t_0)\equiv|\alpha\rangle\langle\alpha|$.
Fig.~\ref{trace} shows an example of their evaluation with
$\Omega=3$ meV, $\kappa=1$ ns$^{-1}$ and $\varepsilon=0.1$.

\begin{figure}[h]
\begin{center}
\epsfig{figure=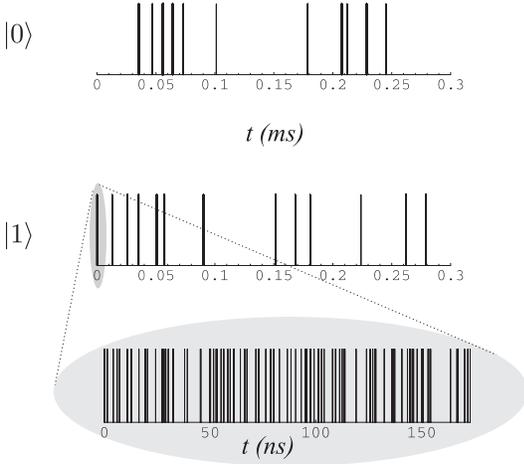,width=7truecm}
\caption{\label{counts}Simulation of photon counts for a system
starting from state $|0\rangle$ (top) and from state $|1\rangle$
(middle). An expanded view of the first few photon counts is
displayed in the bottom graph. Parameters are the same as in Fig.
\protect\ref{trace}.}
\end{center}
\end{figure}

Note that, in contrast to the ``common'' lambda configuration
\cite{Zoller87}, in Eq.~(\ref{eqrho}) both of the recycling terms,
$\kappa\tilde{\rho}_{xx}$ and
$\varepsilon^2\kappa\tilde{\rho}_{xx}$, are missing, since it is
the same photon that induces both these transitions, i.e. we can
not distinguish between the two transitions via photon detection.
This implies that, when the first photon is emitted, say at time
$t_1$, the system collapses either into state $|0\rangle$ --~with
probability $p_0=\varepsilon^2/(1+\varepsilon^2)$~-- or into state
$|1\rangle$ --~with probability $p_1=1/(1+\varepsilon^2)$~--,
whence the evolution starts over again. Therefore the probability
that, at the time $t>t_i$ ($i\geq 1$), the $(i+1)$-th photon has
not been emitted, is
\begin{equation}
\label{Pi} P^{(i)}_\alpha(t-t_i)=\frac{\varepsilon^2
P^{(0)}_0(t-t_i)+P^{(0)}_1(t-t_i)}{1+\varepsilon^2},
\end{equation}
which is independent of the initial state $|\alpha\rangle$. A
typical photoemission pattern will look like Fig.~\ref{counts}: a
sequence of pulses, each one made out of a bunch of the order of
$1/\varepsilon^2$ photons, separated by no-emission windows. This
is the typical quantum-jump pattern one obtains in the presence of
an emission probability having the form of a sum of different
exponentials like Eq.~(\ref{Pi}).

The only feature which allows for discriminating the two patterns
is the first bunch of photons, which are emitted almost
immediately in the case of state $|1\rangle$, and after a sensible
delay in the case of state $|0\rangle$, due to the fact that,
prior to the first photoemission, it was still
$P^{(0)}_0(t)\not=P^{(0)}_1(t)$. Therefore, a detector with 100\%
efficiency would be still capable of discriminating between the
two logical states even in the presence of hole mixing.

Another option would be available in II-VI semiconductor systems,
showing energetic inversion between light and heavy hole states as
described in Sect.~\ref{sec:singlequbit}. To be specific, let us
refer to the left part of Fig.~\ref{singlequbit}. In that case,
the transition to be excited for probing the QD state is the one
marked as (2). Hole mixing results in an unwanted coupling to the
transition (1). This can be compensated for by simply adding a
small component of $\sigma_+$ light, proportional to the mixing
parameter $\varepsilon$. The error affecting this operation would
scale linearly with the imprecision in $\varepsilon$, which is not
straightforward to predict theoretically, yet it can be measured
in a real situation to a good accuracy.

\subsection{Case of perfect detection} We start by
considering the error for the case $\eta=1$ where $\eta$ is the
parameter describing the detection efficiency. On introducing hole
mixing the detector would still be able to discriminate between
the two logical states but mixing will be the cause for two types
of measurement errors which can occur. Starting with the system in
state $|1 \rangle$ there is a possibility for no photon to be
emitted from the QD during the whole measurement time. The
probability for this type of error is given by: $P^{(0)}_1(t)$. In
the other case starting with the system initially in $|0 \rangle$
at least one photon might be emitted during the measurement time.
The probability for this sort of error is given by:
$1-P^{(0)}_0(t)$. The measurement time has to be chosen in such a
way as to minimize the sum of these two errors. For the same
parameters employed in Fig.~\protect\ref{trace}, we obtain an
estimate for the optimal measurement time of the order of a few
tens of ns. What typically happens in practice is that, as shown
in Fig.~\ref{counts}, by appropriate time windowing the first
bunch of photons coming from state $|1\rangle$ can be safely
discriminated from the (later) photons coming from state
$|0\rangle$.

\subsection{Finite detection efficiency} We now consider the case
in which $\eta <1$. The lowest detector efficiency in which we can
still hope to discriminate between the two logical states is given
by $\eta =1 / \langle N \rangle$, where \cite{remark} \ $\langle N
\rangle=1/\varepsilon^2$ is the average number of photons to be
emitted before a system starting off in state $|1\rangle$ flips to
state $|0\rangle$. In our case this is not a tight constraint,
since semiconductor photo-detectors have a very high quantum
efficiency \cite{Bachor}  $\eta_{mat} \approx 0.98 $. The typical
wave length emitted by the recombination process in the QDs lies
well within the spectral window which is due to the cutoff by band
gap energy of such detectors. The main source for low detection
efficiency is due to the probability for the emitted photon to
reach the detector, i.e. the difficulty arising due to finite
angle coverage of the detector. The situation however can be
significantly improved by coupling the QD with a microcavity as
described in \cite{Imamoglu00}.

It is important to note that we can use an avalanche photon
counting mode so that each photon arriving creates one e-h pair
which then amplifies in the device to produce a current spike. The
need to wait a few nanoseconds before detecting the next photon is
not a limitation in our case since the measurement process we are
considering is essentially a one shot measurement process, as long
as the dark count is low enough.

Working with a detector with a finite efficiency means that we
have to choose the measurement time so as to ensure the
fluorescent state emits a few photons thus increasing the
probability one of them will be detected. This increases the
probability for an error due to a photon being emitted by the
initial state $|0 \rangle$ since $P^{(0)}_0(t)$ decays
exponentially with time. Moreover a further possibility for error
is introduced into our measurement scheme. Starting initially in
state $|1 \rangle $ the QD can emit a photon/photons which will go
undetected and the spin can flip into state $|0 \rangle$, i.e.,
the information regarding the spin state is lost without being
detected. The probability $P_e$ for such an error is given by
\begin{eqnarray}
P_e&=&\varepsilon^2\sum_{n=0}^{N} \left(\frac{1-
\eta}{1+\varepsilon^2}
\right)^{n+1}\nonumber\\
&=&\frac{\varepsilon^2 (1- \eta)}{\varepsilon^2 +\eta}\left [1-
{\left ({1- \eta \over 1+ \varepsilon^2}\right )}^{N+1}\right ],
\end{eqnarray}
which is simply the sum over $n$ incidents in which the emitted
photons were not detected and no spin-flip occurred and on the
$n+1$ incident such a spin flip occurred (without the photon being
detected). This type of error turns out not to be particularly
sensitive to the time of measurement. Given that the number of
photon emitted in the first bunch in Fig.~\ref{counts} is of the
order of $\varepsilon^{-2}\approx 10^{2}$ in our case as discussed
above, and taking an efficiency $\eta=0.9$ we obtain an error due
to finite detection efficiency of the order of 0.1\%.

\section{Conclusions}

To claim that a certain implementation scheme for quantum
information processing is viable, one has to carefully understand
the fundamental sources of decoherence acting in that specific
physical system, and to show that they can actually be controlled.
To this aim, in this paper we analyzed in detail the different
decoherence mechanisms affecting a recently proposed all-optical
scheme for quantum computation based on electron spin in quantum
dots. In particular, we took into account the effect of hole
mixing and of coupling to phonons at a finite temperature,
estimating their impact on each of the building blocks of a
quantum computer: single- and two-qubit gates, and state read-out.
We developed a strategy to circumvent such unwanted effects via an
adiabatic laser excitation scheme, simulated its performance under
realistic conditions and evaluated the corresponding fidelity. Our
scheme turns out to be able to suppress the effect of both of
these decoherence sources on the proposed gate, and therefore
constitutes a viable proposal for all-optical quantum information
processing in semiconductor quantum dots.

\acknowledgments

This work was performed with partial support from the EC project
CECQDM (IST-2001-38950). T.C. acknowledges support from the
Istituto Trentino di Cultura, the Italian Fulbright Commission and
the National Institute of Standards and Technology; A.D. thanks
the Institute of Theoretical Physics of Innsbruck University for
hospitality and support.

\appendix

\section{Coupling constants: Microscopic calculation\label{sec:Coupling-constants:-Microscopic}}

The phonons are coupled to the charge distribution in a QD by
means of either deformation or piezoelectric coupling potentials.
The corresponding coupling constants are either \begin{equation}
\lambda _{q}^{D}=\frac{qD(q)}{\sqrt{2\rho \omega
(\mathbf{q})V}}\label{eq:inilamD}\end{equation} in the case of
deformation coupling, or \begin{equation} \lambda
_{\mathbf{q}}^{p}=\frac{\rho (\mathbf{q})}{\sqrt{2\rho \omega
(\mathbf{q})V}}M(\mathbf{q}),\label{eq:inilamP}\end{equation} in
the case of piezoelectric coupling. In both cases $\rho $ is the
mass density of the sample, and\begin{equation}
M(\mathbf{q})=\frac{24\pi ee_{14}}{\epsilon
q^{3}}q_{x}q_{y}q_{z}\equiv
M\frac{q_{x}q_{y}q_{z}}{q^{3}},\end{equation} is the coupling
potential. Here $\epsilon $ is the dielectric constant of the
sample and $e_{14}$ is the material constant. The form-factors
\begin{equation}
\rho (\mathbf{q})=\int d\mathbf{r}\big[|\psi
_{v}(\mathbf{r})|^{2}-|\psi _{c}(\mathbf{r})|^{2}\big]\exp
(-i\mathbf{q}\mathbf{r}),\end{equation} and \cite{takagahara}
\begin{equation} D(\mathbf{q})=\int d\mathbf{r}\big[D_{v}|\psi
_{v}(\mathbf{r})|^{2}-D_{c}|\psi _{c}(\mathbf{r})|^{2}\big]\exp
(-i\mathbf{q}\mathbf{r})\end{equation} are related to the exciton
charge density. The wavefunctions $\psi _{v}$and $\psi _{c}$
describe the hole and the electron states making up the exciton.
$D_{c},D_{v}$ are the deformation coupling potentials.

The calculation of the coupling constants relies on a microscopic
model. We consider a QD in a static external electric field $F_0$
directed along the $x$-axis. Within the simplest model with
harmonic confinement potential the Hamiltonian for the
single-particle electron ($i=e$) and the holes ($i=h$) states is
given by \begin{equation} H=\frac{p^{2}}{2m_{i}}+\frac{m_{i}\omega
_{i}^{2}(r+r_{0i})^{2}}{2}-\frac{m_{i}\omega
_{i}^{2}r_{0i}^{2}}{2},\label{eq:dotHam}\end{equation} where
$r_{0i}=e_{i}F_0/m_{i}\omega _{i}^{2}$ is a measure of the
electric field strength in {}``oscillator units of length'',
$\omega _{i}$ is the frequency of the confining potential, $m_{i}$
and $e_{i}$ are the mass and the charge of the particles (the
electrons or holes).

To clarify the effects of external electric field and compare the
relative strength of the different types of the coupling, let us
consider first the somewhat unrealistic but otherwise simple model
of a spherically symmetric QD. In this case the wavefunction is
given by\begin{equation} \psi _{i}=\left(\frac{1}{\pi
l_{i}^{2}}\right)^{3/2}\exp
\left\{-\frac{(\mathbf{r}-\mathbf{r}_{0i})^{2}}{2l_{i}^{2}}\right\},\end{equation}
where $l_{i}=(m_{i}\omega _{i})^{-1/2}$ is the ground state
localization length. Then, in the case of deformation coupling, a
simple calculation gives the following expression for $J-$function
(for simplicity we put $l_{e}=l_{h}$):\begin{eqnarray} J(\omega
)&=&\frac{\omega ^{3}}{4\pi ^{2}\rho
u^{5}}\exp\Bigl\{-\frac{\omega
^{2}l^{2}}{2u^{2}}\Bigr\}\nonumber\\
&&\mbox{}\times\Bigl[D_{e}^{2}+D_{h}^{2}-2D_{e}D_{h}\frac{\sin
(2\omega r_{0}/u)}{2\omega r_{0}/u}\Bigr].\end{eqnarray} The
obtained result shows the two important features. First of all,
since $D_{e}\neq D_{h}$ the $J-$function is superohmic in the
external field of any strength. Essentially this means that a
static electric field does not qualitatively change the
interaction with the phonons. Secondly, the function $J(\omega )$
has a large frequency cutoff at $\omega _{l}=u/l$, which is
nothing else but the inverse flight time of a phonon through the
QD.

The piezoelectric coupling behaves somewhat different.
Recalculating the $J-$function using the coupling constant
(\ref{eq:inilamP}) we find
\begin{eqnarray} J^{p}(\omega
)&=&\frac{M^{2}\omega }{560\pi ^{2}\rho u^{3}}\Bigl[\exp
\Bigl(-\frac{\omega ^{2}l_{c}^{2}}{2u^{2}}\Bigr)+\exp
\Bigl(-\frac{\omega ^{2}l_{v}^{2}}{2u^{2}}\Bigr)\nonumber\\&
-&2\exp \Bigl(-\frac{\omega ^{2}l_{c}^{2}}{4u^{2}}\Bigr)\exp
\Bigl(-\frac{\omega
^{2}l_{v}^{2}}{4u^{2}}\Bigr)f\Bigl(2\frac{\omega
}{u}r_{0}\Bigr)\Bigr]\label{eq:JPE}\end{eqnarray} where $f(0)=1$
and $f(x)\rightarrow 0$ at $x\rightarrow \infty $. Since the
piezoelectric potential $M$ is the same for the electrons and for
the holes, in the limit of small electric field strength we
have:\begin{equation} J_{F_0=0}^{p}(\omega \rightarrow
0)=\frac{M^{2}\omega ^{5}(l_{c}^{2}-l_{v}^{2})^{2}}{6720\pi
^{2}\rho u^{7}}.\end{equation} The coupling is still superohmic,
but it contains a larger power of $\omega $ than the deformation
coupling. Moreover, the value of the coupling potential $M$ is
also numerically small for common materials and thus the
interaction of a QD with the phonons is dominated by the
deformation coupling.

The large electric field limit of Eq. (\ref{eq:JPE}) is somewhat
interesting. One can see that we have \begin{equation}
J_{F_0=\infty }^{p}(\omega \rightarrow 0)=\frac{M^{2}\omega
}{280\pi ^{2}\rho u^{3}},\end{equation} which is, at first glance,
an indication of ohmic type of coupling. Nevertheless, one should
keep in mind that the $J^{p}-$function decreases quickly when
$\omega \agt \omega _{l}$ and hence the argument of the
$f$-function never exceeds $\sim r_{0}/l$. This means that the
function $f\sim 1$ practically everywhere in the course of the
integration in Eq. (\ref{eq:Jdef}) and hence the coupling remains
superohmic in the whole relevant parameters range.

The analysis above paves the way to a more realistic calculation.
Consider the Hamiltonian (\ref{eq:dotHam}) acting in 2D ($x$- and
$y$- directions), whereas the motion in the $z$-direction is
confined within a box of length $L_{z}$. The wavefunction (of the
ground state) is \begin{equation} \psi
_{i}=\sqrt{\frac{2}{L_{z}}}\sin \frac{\pi z}{L_{z}}
\sqrt{\frac{m_{i}\omega _{i}}{\pi }}\exp
\Bigl\{-\frac{(r+r_{0i})^{2}}{2l_{i}^{2}}\Bigr\},\end{equation}
where $l_{i}=\sqrt{m_{i}\omega _{i}}$. The Fourier component is
given by\begin{equation} \int d^{3}r\psi ^{2}\exp
\{-i\mathbf{q}\mathbf{r}\}=F_{z}(q_{z})F(\mathbf{q}),\end{equation}
where $\mathbf{q}=(q_{x},q_{y})$ is the vector in the $xy-$plane,
and
\begin{equation}
F(\mathbf{q})=\exp
\Bigl\{-\frac{q^{2}l^{2}}{4}+i\mathbf{q}\mathbf{r}_{0}\Bigr\},\end{equation}
and
\begin{equation} F_{z}(q_{z})=\frac{4\pi ^{2}i[\exp
(iq_{z}L)-1]}{(q_{z}L)^{3}-4\pi ^{2}q_{z}L}.\end{equation} We note
that, in spite of its ugly appearance, the form-factor $F_{z}$ is
nowhere singular on the real axis and quickly decays when
$q_{z}L\gg 1$. The normalization ensures that $F_{z}(0)=1$.

For simplicity consider the limit of very strong confinement:
$L_{z}\ll l_{c,v}$. Then, neglecting the piezoelectric coupling
and using the zero-argument value for the function $F_{z}$, we
obtain\begin{equation} J(\omega )=\frac{\omega ^{3}}{4\pi ^{2}\rho
u^{5}}\exp \Bigl\{-\frac{\omega
^{2}l^{2}}{2u^{2}}\Bigr\}\Bigl[D_{c}^{2}+D_{v}^{2}-2D_{c}D_{v}f_{1}\Bigl(2\frac{\omega
}{u}r_{0}\Bigr)\Bigr],\end{equation} where we have
\begin{equation} f_{1}(x)= \frac{1}{4\pi }\int d\theta\sin\theta\;
d\phi\;\cos (x\sin \theta \cos \phi ).\end{equation} At small
value of its argument this function gives $1$ and vanishes when
$x\rightarrow \infty $. The presented $J-$function is always
superohmic and is not qualitatively different from that considered
above for an idealistic spherically symmetric QD. It shares all
the important features of the simpler model above. In particular,
the large-frequency cutoff is defined by the same inverse phonon
flight-time through the QD: $\omega _{l}\sim u/l$.

In short, we presented a number of examples of the spectral
function calculations. Both in the case of piezoelectric and of
deformation coupling, the $J$-function is superohmic, with the
exponents $s=3$ and $s=5$, respectively. The quantity $\omega
_{l}\sim u/l$ plays the role of the high frequency cutoff.

\section{Adiabatic effective Hamiltonian: Applicability of perturbation expansions\label{sec:Adiabatic-effective-Hamiltonian.}}

The phonon interaction terms in the Hamiltonian (\ref{eq:iniHam})
may well be large and hence the perturbation theory expansion in
powers of $\lambda _{\mathbf{q}}$ may not always be well
justified. A remarkable opportunity to extract non-perturbative
results originates from our adiabatic assumption.

Indeed our quantum gate proposal relies on adiabatic manipulations
of the external parameters of the Hamiltonian (\ref{eq:iniHam}).
Assume that both $\Omega (t)$ and $\Delta (t)$ change on a time
scale $\tau $. Extreme adiabatic condition implies
\begin{equation} \omega _{l}\tau \gg
1,\label{eq:adiabextreme}\end{equation} i.e. the considered QD
dynamics can be considered slow for almost all of the phonon
modes. According to \cite{Leggett}, this condition can be formally
used to adiabatically eliminate all the phonon modes with
frequencies exceeding a certain cutoff $\omega _{*}$: $\tau
^{-1}\ll \omega _{*}\ll \omega _{l}$ and obtain the effective
adiabatic Hamiltonian for the slow phonon modes with frequencies
$\omega \alt \omega _{*}$. At zero temperature $T=0$ the effective
Hamlitonian takes the form:\begin{eqnarray} H_{\rm
ph}&=&\Bigl[-\tilde{\Delta }+\sum _{\mathbf{q},\omega
(\mathbf{q})<\omega _{*}}\lambda
_{\mathbf{q}}(b_{\mathbf{q}}+b_{\mathbf{q}}^{\dagger
})\Bigr]|e\rangle \langle e|\label{eq:adiaHam}\\&&
+\frac{\tilde{\Omega }}{2}(|e\rangle \langle g|+|g\rangle \langle
e|)+\sum _{\mathbf{q},\omega (\mathbf{q})<\omega _{*}}\omega
(\mathbf{q})b_{\mathbf{q}}^{\dagger
}b_{\mathbf{q}}.\nonumber\end{eqnarray} The Hamiltonian
(\ref{eq:adiaHam}) has the same form as the original Hamiltonian
(\ref{eq:iniHam}). The effect of the high frequency mode is
limited to the renormalization of the frequency
detuning\begin{equation} \tilde{\Delta }=\Delta -\frac{1}{2}\int
d\omega \frac{J(\omega )}{\omega
},\label{eq:deltaren}\end{equation} and Rabi frequency (tunneling
term in the spin boson model) \begin{equation} \tilde{\Omega
}=\Omega \exp \left(-\int _{\omega _{*}}^{\infty }\frac{d\omega
}{2}\frac{J(\omega )}{\omega
^{2}}\right),\label{eq:omegaren}\end{equation}
 In the superohmic case the integral in the exponent of Eq. (\ref{eq:omegaren})
converges at $\omega \alt \omega _{*}$ and hence does not depend
on $\omega _{*}$. Thus, in the adiabatic approximation we can
conveniently set $\omega _{*}=0$. This means that at zero
temperature even a strong phonon coupling leads only to the
renormalization of the Hamiltonian parameters (the Rabi
frequency). The measure of the Rabi frequency renormalization
yields the perturbation theory expansion
parameter:\begin{equation} \frac{\Omega -\tilde{\Omega }}{\Omega
}\sim \int \frac{d\omega }{2}\frac{J(\omega )}{\omega ^{2}}\sim
\frac{J(\omega _{l})}{\omega _{l}}\ll
1.\label{eq:expanparam}\end{equation} The effective Hamiltonian
(\ref{eq:adiaHam}) does not rely on this condition, whereas a
perturbation theory for arbitrary fast processes would do.

The situation is somewhat different at finite temperatures. The QD
can still change its state {}``coherently'', i.e. in the course of
Rabi oscillations which are characterized by the new
temperature-dependent renormalized value, called Huang-Rhys factor
in \cite{Leggett}:\begin{equation} \tilde{\Omega }=\Omega \exp
\Bigl\{-\int _{0}^{\infty }\frac{d\omega }{2}\frac{J(\omega
)}{\omega ^{2}}[1+2N(\omega
)]\Bigr\},\label{eq:omegarenT}\end{equation} where $N(x)=[\exp
(x/T)-1]^{-1}$ is the Bose occupation number. This is nothing else
but the generalization of Eq. (\ref{eq:omegaren}) now taking into
account finite occupation of the phonon modes. In addition to
that, the QD can change its state by either absorbing or emitting
a phonon. This process is called incoherent tunnelling and has no
effective Hamiltonian form (generally speaking there is no obvious
separation between the fast and slow variables).

In the quantum gate proposal we suggest to operate our qubit
starting from the ground state and adiabatically changing
parameters. This means that the QD cannot emit a phonon of a high
energy $\omega \agt \tau ^{-1}$ (since it is in the ground state).
The absorption of a phonon from the thermal bath requires finite
occupancy of a state with energy $\omega \sim \tilde{\Omega }$,
which can be made exponentially small, provided $T\ll
\tilde{\Omega }$.

The results of this section also make sense if compared with
perturbation theory. Consider the case of the phonon coupling.
There is no corrections to the eigenenergies (\ref{eq:adiaen}) to
first order in powers of $\lambda _{q}$. The second order
perturbation theory gives \begin{equation} \delta E_{\pm }=\sum
_{q}\frac{\lambda _{q}^{2}}{4}\left\{\frac{[1\pm \cos (\theta
)]^{2}}{\omega _{q}}+\frac{\sin ^{2}(\theta )}{\sqrt{\Delta
^{2}+\Omega ^{2}}+\omega _{q}}\right\}.\end{equation} In the
adiabatic limit (\ref{eq:adiabextreme}) we can expand in powers of
$\Delta /\omega _{l}$ and $\Omega /\omega _{l}$ to obtain the
expression:\begin{eqnarray} \delta E_{\pm }&=&\frac{1}{2}(1\pm
\cos \theta )\int d\omega \frac{J(\omega )}{\omega
}\nonumber\\&&\mbox{}-\frac{\sin ^{2}\theta }{4}\sqrt{\Delta
^{2}+\Omega ^{2}}\int d\omega \frac{J(\omega )}{\omega
^{2}}.\end{eqnarray} The same expression could be obtained by
substituting the renormalized values
(\ref{eq:deltaren}-\ref{eq:omegaren}) into the adiabatic energies
(\ref{eq:adiaen}) and expanding the obtained expession in powers
of the small parameter (\ref{eq:expanparam}). This once again
shows that the results of the adiabatic renormalization coincide
with perturbation theory whenever both approaches are equally
valid.

We conclude that the effective Hamiltonian (\ref{eq:adiaHam}) with
renormalized value of the Rabi transition amplitude
(\ref{eq:omegarenT}) can be used as a good non-perturbative tool
to study the QD dynamics in the presence of phonons. Its validity
is ensured by the fact that the $J$-function for the deformation
coupling is superohmic and relies on the timescale separation
condtion (\ref{eq:adiabextreme}). In the case when $\omega
_{l}\tau \agt 1$, the adiabatic approximation fails and one has to
resort to perturbation theory, whose expansion parameter is given
by (\ref{eq:expanparam}).


\begin{thebibliography}{99}
\bibitem{remark}
The value of $\langle N \rangle$ is given by $\langle N \rangle=
\sum_{n=0}^{\infty} n q^n p= q {\partial \over \partial q}
\sum_{n=0}^{\infty} q^n p$, where $q=1/(1+\varepsilon)$ and
$p=1-q$.
\end{thebibliography}
\end{document}